\input pipi.sty
\input epsf.sty
\input psfig.sty
\magnification1000
\raggedbottom

\nopagenumbers
\rightline\timestamp 

\smallskip
\rightline{FTUAM 03-14}
\rightline{hep-ph/0310206}

\bigskip
\hrule height .3mm
\vskip.6cm
\centerline{{\bigfib Comments on some chiral-dispersive calculations
 of $\pi\pi$ scattering}\footnote{*}{\petit This note contains the material 
presented in talks given at 
``Photon 2003" (Frascati, April 2003), ``QCD at Work" 
(Conversano, June 2003),  ``Chiral Dymamics"  (Bonn, September 2003), 
Benasque (July, 2004) and ``Quark Confinement and the Hadron Spectrum"
(Sardinia, September 2004).}}
\medskip
\centerline{\medfib Final, completely revised and improved version}
\medskip
\centerrule{.7cm}
\vskip1.5truecm
\setbox9=\vbox{\hsize65mm {\noindent\fib F. J. Yndur\'ain} 
\vskip .1cm
\noindent{\addressfont Departamento de F\'{\i}sica Te\'orica, C-XI\hb
 Universidad Aut\'onoma de Madrid,\hb
 Canto Blanco,\hb
E-28049, Madrid, Spain.}\hb}
\smallskip
\centerline{\box9}
\vskip2truecm

\setbox0=\vbox{\abstracttype{Abstract}We consider the 
 $\pi\pi$ scattering amplitude obtained by  Colangelo, Gasser and Leutwyler (CGL), 
using dispersion relations and chiral perturbation theory to two loops. 
We point out that the input used by CGL for energies above 1.42~GeV  is   
incompatible with a number of experimental results. 
Moreover, the error they adscribe to the $\delta_0^{(0)}$ phase at 0.8~GeV, an important part
of their input,  is much underestimated, and its central value is probably displaced. 
Then, we compare CGL results with  direct fits to experimental $\pi\pi$ data; 
the outcome of this test is mismatch of the CGL amplitude, for
several quantities  by as much as 4 standard deviations. We also 
evaluate forward dispersion relations with the CGL amplitudes. 
We find that these dispersion relations are fulfilled less and less 
well as the energy increases towards 0.8~GeV. Moreover, the size that the experimental results
for $\pi\pi$ scattering indicates for the two loop corrections in a chiral 
perturbation theory analysis is less than those required by CGL,
 which  suggests that at least some of  the CGL corrections are so large to cover biases in
 their $\pi\pi$  amplitude and in the evaluation of the scalar form factor of the pion.
 We conclude on the necessity of repeating the analysis of CGL with correct high energy 
amplitudes and a more realistic input for the $\delta_0^{(0)}$ phase at 0.8~GeV.}
\centerline{\box0}
\brochureendcover{Typeset with \physmatex}
\brochureb{\smallsc f. j.  yndur\'ain}{\smallsc 
comments on some chiral-dispersive calculations
 of $\pi\pi$ scattering}{1}

\brochuresection{1. Introduction}

\noindent
In two  recent papers, Ananthanarayan, Colangelo, Gasser and Leutwyler\ref{1} 
and Colangelo, Gasser and Leutwyler\ref{2} (to be referred to as, respectively, 
ACGL and CGL) have used 
experimental information, analyticity and unitarity (in the form of  
the Roy equations) and, in CGL, chiral perturbation theory, 
to construct the  $\pi\pi$ 
scattering amplitude at low energy.

We will mainly discuss  the second paper, CGL. 
Here it is given a set of low energy parameters 
(scattering lengths and effective ranges) for S, P, D and F waves
in which an outstanding precision (at the percent level) is claimed. 
For the S0, S2 and P waves\fnote{We use the convenient and
 self-explanatory notation of S0, S2, P, D0, D2, F, 
\tdots, 
for the $\pi\pi$ partial waves.} 
these authors also give  a parametrization for the phase shifts valid for
energies  up to $s^{1/2}=0.8\,\gev$. 

While the results of CGL constitute a substantial improvement over previous ones, 
the analysis of these authors presents  
 a number of drawbacks.  First of all,  the input scattering amplitude 
 which ACGL, CGL use for high energy ($s^{1/2}\geq1.42\,\gev$)
 is not physically acceptable, as will be shown below; and 
 the errors these authors take 
for some of their experimental input data are 
excessively optimistic. 
Finally, the input D2 wave used is incompatible with a number of requirements.
Because of this,  
the CGL threshold parameters, and the low energy S0, S2 and 
P waves below 0.8
\gev, are somewhat displaced from what one gets by direct fits to experimental  data,
 and fail to pass a number of
consistency tests; notably, they do not fulfill well forward 
dispersion relations. All this is discussed in the present note,
 which is based on  recent   
 papers by the author and J.~R.~Pel\'aez, refs.~3,~4 and 5.

\booksection{2. The  partial waves at low energy, $s^{1/2}\leq 1.42\,\gev$, from fits to
data}

\noindent
We will 
first consider  wave-by-wave fits to {\sl experimental} data for the S0, S2, P waves, as 
given in ref.~5. This improves the 
 set of phase shifts, 
which was called ``tentative solution"  in ref.~3, 
and will provide us with a handy representation of what experiment gives 
for the $\pi\pi$ scattering amplitudes. 
To fit the various phase shifts, $\delta_l^{(I)}(s)$, we use a parametrization for 
 $\cot\delta_l^{(I)}(s)$ which takes into account the analyticity properties 
of this quantity, as well as its zeros (associated with resonances) 
and poles (when the phase shift crosses $n\pi$, $n=$integer).
Explicit expressions will be given in each specific case.

\booksubsection{2.1. The S0, S2 and P partial waves at low energy, $s^{1/2}\lsim 1\,\gev$}

\noindent
We start with  the P wave, for which
 we take the results obtained from a fit to the pion form
factor, both in the spacelike and timelike  regions, 
as given in ref.~6. 
We write
$$\cot\delta_1(s)=\dfrac{s^{1/2}}{2k^3}
(M^2_\rho-s)\left\{B_0+B_1\dfrac{\sqrt{s}-\sqrt{s_0-s}}{\sqrt{s}+\sqrt{s_0-s}}
\right\};\quad s_0^{1/2}=1.05\;\gev.
\equn{(2.1a)}$$
$s_0$ is the point at which inelasticity begins to be nonegligible.
We find
$$\eqalign{
B_0=&\,1.069\pm0.011,\quad B_1=0.13\pm0.05,\quad M_{\rho}=773.6\pm0.9,\cr 
a_1=&\,(37.6\pm1.1)\times10^{-3}M_{\pi}^{-3},\quad
  b_1=(4.73\pm0.26)\times10^{-3}M_{\pi}^{-5};
}
\equn{(2.1)}$$
this  parametrization 
 we take to be valid up to $s^{1/2}=1\,\gev$. 
The fit to $\pi\pi$ data is shown in \fig~1.

\topinsert{
\setbox0=\vbox{{\psfig{figure=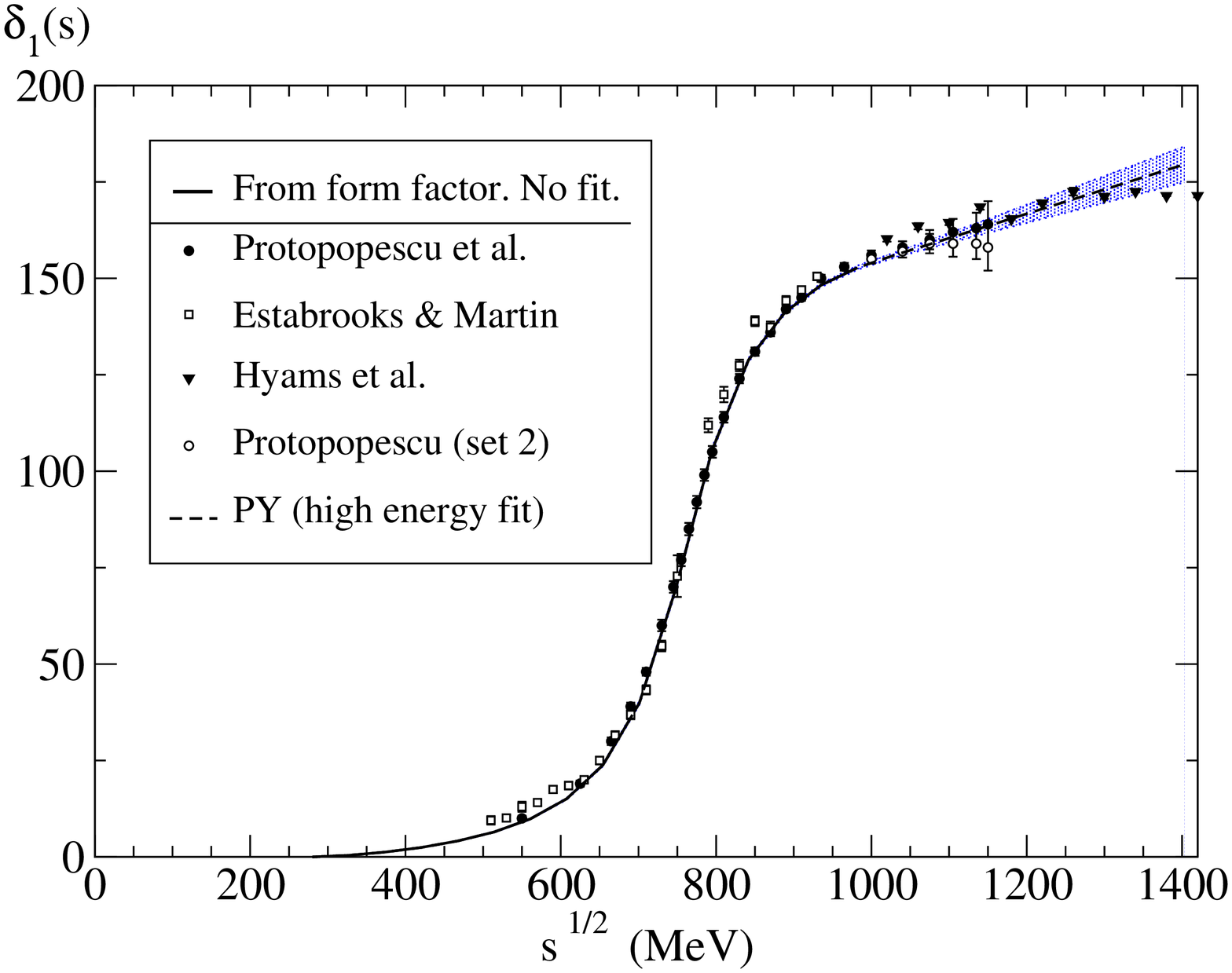,width=11.2truecm,angle=-90}}}
\setbox6=\vbox{\hsize 15truecm\captiontype\figurasc{Figure 1. }{The  
phase shifts of solution 1 from Protopopescu et al.,\ref{9} 
Hyams et al.\ref{10a} and 
Estabrooks and Martin\ref{10c} compared with the prediction with 
the parameters (2.1)   (solid line below 1~\gev). 
We emphasize that this is {\sl not} a fit 
to these experimental data, but is obtained 
from the pion form factor.\ref{6} The error here is like the thickness of the line.
\hb
\quad Above 1~\gev, the dotted line and error (PY) are as 
follows from the fit in ref.~5, \subsect~2.1.2.}
} 
\medskip
\centerline{\tightboxit{\box0}}
\bigskip
\centerline{\box6}
\bigskip
}\endinsert
  
For the S2 wave we fit data where two like charge pions are produced:\ref{7} 
although these pions are not all on their mass shell, at least there is no 
problem of interference among various isospin states. 
In the low energy region,  we  fix the Adler zero at $z_2=M_\pi$ and
 fit only the low energy data, 
$s^{1/2}<1.0\,\gev$; later on we will allow $z_2$ to vary. We 
have
$$\cot\delta_0^{(2)}(s)=\dfrac{s^{1/2}}{2k}\,\dfrac{M_{\pi}^2}{s-2z_2^2}\,
\left\{B_0+B_1\dfrac{\sqrt{s}-\sqrt{s{_0}-s}}{\sqrt{s}+\sqrt{s{_0}-s}}\right\},
\quad z_2\equiv M_\pi;\quad s{_0}^{1/2}=1.05\;\gev,
\equn{(2.2a)}$$  
$$\eqalign{
B_0=&\,-80.4\pm2.8,\quad B_1=-73.6\pm12.6;\cr
a_0^{(2)}=&\,(-0.052\pm0.012)\,M_{\pi}^{-1};\quad
b_0^{(2)}=(-0.085\pm0.011)\,M_{\pi}^{-3}.\cr
}\equn{(2.2b)}$$

 For S0, where the experimental situation is somewhat 
confused, we consider two different methods of data selection. 
In both, we fit those 
experimental data points in which  {\sl on mass shell} pions are 
involved: $K_{l4}$ and $K\to2\pi$ decays. In the first method, we also
 include some points at higher 
energy ($0.81\,\gev\leq s^{1/2}\leq0.97\,\gev$) where (most of) 
the various experimental
 numbers agree one with
another within 
$\lsim1.5\,\sigma$. This we will call a {\sl global fit}.
Care is exercised to compose the errors in a realistic manner; 
the details  may be found in 
ref.~5, \subsect~2.2.2. 
In this case we fix the Adler zero at $z_0=M_\pi$ and find
$$\eqalign{
\cot\delta_0^{(0)}(s)=&\,\dfrac{s^{1/2}}{2k}\,\dfrac{M_{\pi}^2}{s-\tfrac{1}{2}z_0^2}\,
\dfrac{M^2_\sigma-s}{M^2_\sigma}\,
\left\{B_0+B_1\dfrac{\sqrt{s}-\sqrt{s_0-s}}{\sqrt{s}+\sqrt{s_0-s}}\right\},
\quad z_0\equiv M_\pi;\cr
{B}_0=&\,21.04,\quad {B}_1=6.62,\quad
M_\sigma=782\pm24\,\mev;\quad {\chi^2}/{\rm d.o.f.}={15.7}/(19-3).\cr
\quad 
a_0^{(0)}=&\,(0.230\pm0.010) M_{\pi}^{-1},\quad b_0^{(0)}=(0.268\pm0.011)
M_{\pi}^{-3};\quad\delta_0^{(0)}(m_K)= 41.0\degrees\pm2.1\degrees;
\cr    }
\equn{(2.3a)}$$
this fit (shown in \fig~2, PY curve) we take to be valid for $s^{1/2}\leq0.95\,\gev$. 
The errors of the $B_i$ are strongly correlated; uncorrelated errors are obtained if 
replacing the $B_i$ by the 
parameters $x,\,y$ with
$$B_0=y-x;\quad B_1=6.62-2.59 x;\quad y=21.04\pm0.70,\quad x=0\pm 2.6.
\equn{(2.3b)}$$

\topinsert{
\medskip
\setbox0=\vbox{{\psfig{figure=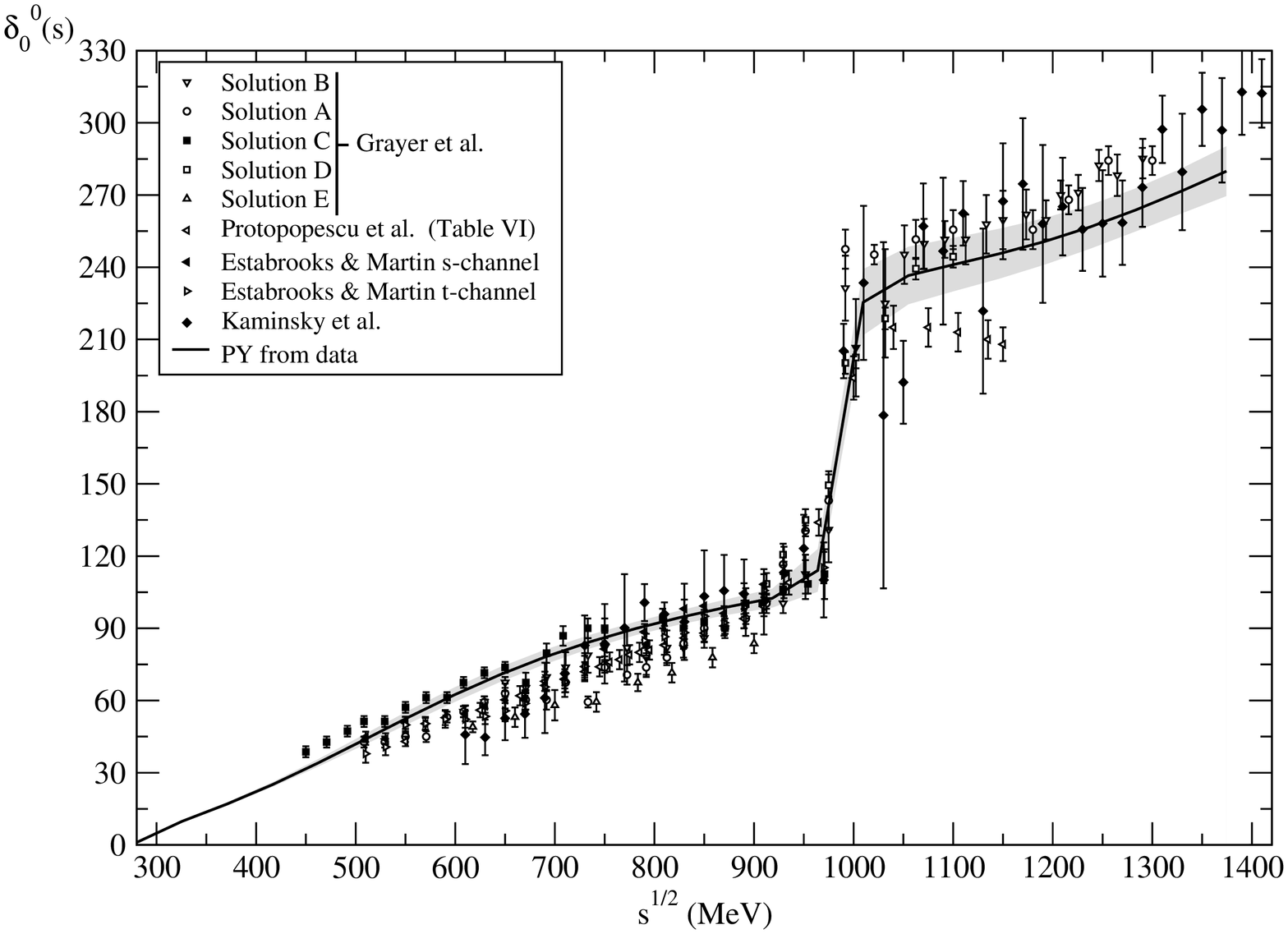,width=14truecm,angle=-90}}}
\setbox6=\vbox{\hsize 15truecm\captiontype\figurasc{Figure 2. }{
The  
S0 phase shifts and error band in the whole energy range as 
given by (2.3) below 1~\gev, and from ref.~5 (\subsect~2.2.4) above 1~\gev.
  The experimental points  from 
$K_{l4}$ and
$K_{2\pi}$ decays are not shown.}} 
\centerline{\tightboxit{\box0}}
\bigskip
\centerline{\box6}
}\endinsert 

Alternate possibilities are to fit only  $K_{l4}$ and $K\to2\pi$ data, or 
to add to this, individually, data from the various experimental analyses.
The results can be found in Table~1 (see below, \subsect~4.1).

\booksubsection{2.2. The S0, S2 and P  waves at intermediate\
 energy, $1\,\gev\lsim s^{1/2}\lsim 1.42\,\gev$}

\noindent
We will not give here details on the fits to experimental data for the 
 S0, S2 and P waves from $\sim1\,\gev$ up to 1.42~\gev, which may be found in ref.~5. 
We will only comment that, while the S2 wave is reasonably well determined from experiment, 
the S0 and P waves are not. Thus, for the first, depending on whether we use 
$\pi\pi\to\pi\pi$ data, or data\ref{11} on 
$\pi\pi\to\bar{K}K$, the inelasticity  $1-\eta_0^{(0)}$
 varies by almost a factor {\sl three}; 
we will discuss this again 
in connection with the scalar form factor of the pion in \sect~9. 
Likewise, the values for $\delta_1(s)$ given in the various 
experimental analyses disagree 
above 1.15~\gev. 
This means that relations, such as dispersion relations, in which the S0, P waves intervene, 
will have errors beyond the nominal errors  above 
1~\gev.

\booksubsection{2.3. The  D, F partial waves at low energy, $s^{1/2}\lsim 1.42\,\gev$,
 from fits to
data}

\noindent
The D and F waves cannot be described with the same accuracy as the S, P waves. 
The data are scanty, and have large errors. 
To get reasonable fits  we will impose the 
values of the scattering lengths that follow from the Froissart--Gribov representation. 
Note that this is not circular reasoning, and it only 
introduces a small correlation: the  Froissart--Gribov representation 
for the D0, D2, F waves depends mostly on the S0, S2 and P waves, and very little on the
 D0, D2, F waves themselves. 
We do not discuss here the D0 and F waves (that may be found in detail  
in ref.~5) as they do not present special features; only the D2 wave will be 
considered now.

\topinsert{
\setbox0=\vbox{{\hfil\epsfxsize 12.5truecm\epsfbox{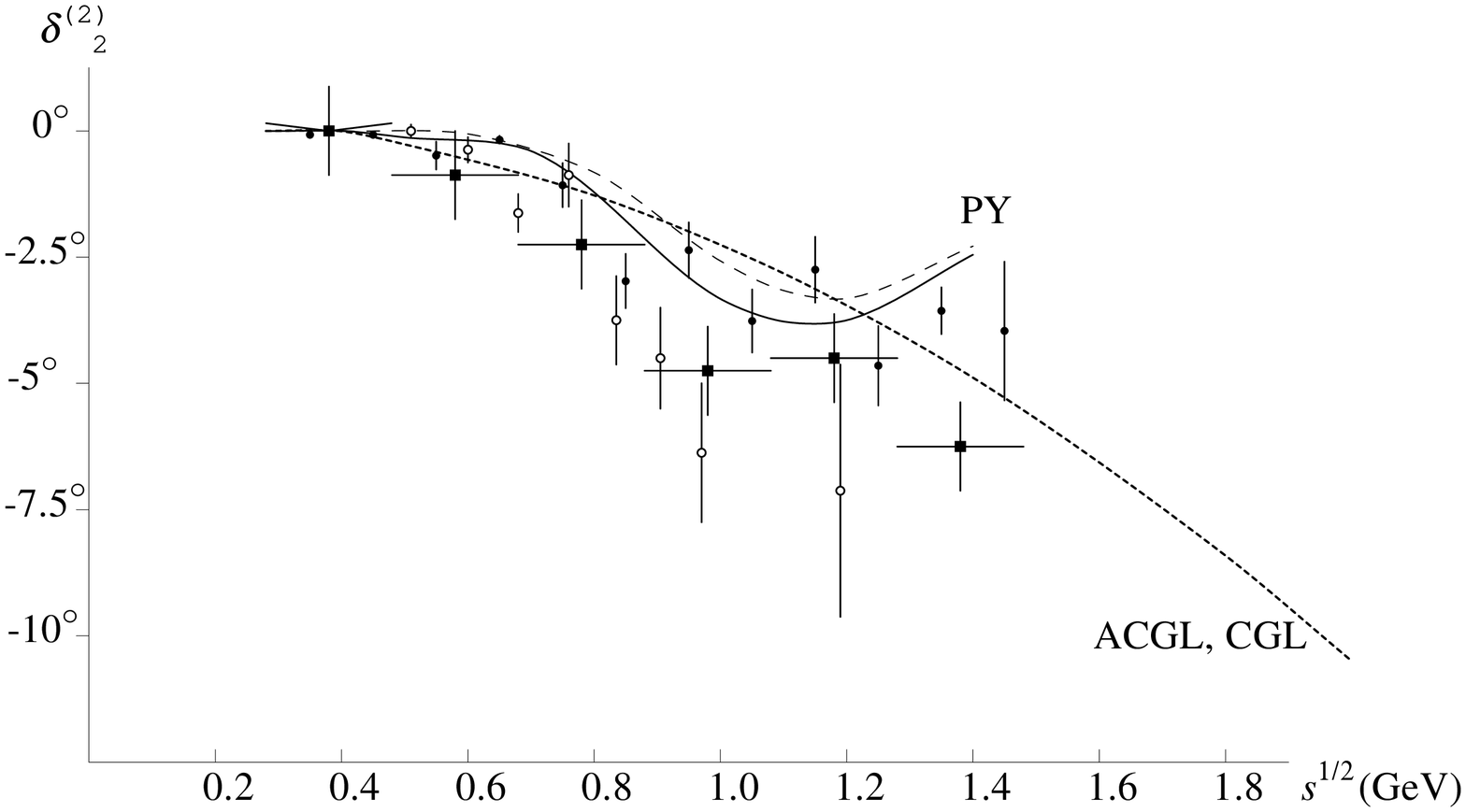}\hfil}}  
\centerline{{\box0}}
\setbox6=\vbox{\hsize 14truecm\captiontype\figurasc{Figure 3. }{
Continuous line: The  
$I=2$, $D$-wave phase shift, obtained by only fitting the 
experimental data. Dashed line: with the parameters improved using 
dispersion relations. Dotted line: the 
fit, valid between $s^{1/2}=0.625\,\gev$ and 1.375~\gev, of 
Martin, Morgan and Shaw which ACGL and CGL, however, use from threshold 
to  $s^{1/2}=2\,\gev$.  
The experimental points are from Losty et al., Hoogland et al.,
 solution A, and Cohen et al. (ref.~7) 
}} 
\centerline{\tightboxit{\box0}}
\centerline{\box6}
\medskip
}\endinsert

For isospin $I=2$ we would only expect important inelasticity when the channel 
 $\pi\pi\to\rho\rho$  opens up, 
so we will take the value $s_0=1.45^2\,\gev^2\sim4M^2_\rho$ for the energy at which 
elasticity is not negligible.
A pole term is necessary here to get an acceptable fit down to low energy,
since we expect $\delta_2^{(2)}$ to change sign near threshold.
The experimental measurements (Cohen et al.; Losty et al.; Hoogland et al.\ref{7}) give 
negative and small values for the phase above some $500\,\mev$, while, 
from   
 the Froissart--Gribov representation, it has been known for a long time\ref{12}
 that the  
scattering length must be positive.
 If we want a parametrization that 
applies down to threshold, we must incorporate this  
zero of the phase shift. 
The inflection seen in data around 1~\gev\ implies that 
we will have to go to third order in the 
conformal expansion. So 
we write
$$\cot\delta_2^{(2)}(s)=
\dfrac{s^{1/2}}{2k^5}\,\Big\{B_0+B_1 w(s)+B_2 w(s)^2\Big\}\,
\dfrac{{M_\pi}^4 s}{4({M_\pi}^2+\deltav^2)-s}
\equn{(2.4a)}$$
with $\deltav$ a free parameter fixing the zero and
$$w(s)=\dfrac{\sqrt{s}-\sqrt{s_0-s}}{\sqrt{s}+\sqrt{s_0-s}},\quad
 s_0^{1/2}=1450\,\mev.$$
Since  the data we have on this wave are not accurate (cf.~\fig~3)  we 
include in the fit the value of  
 the scattering length  that follows
 from the Froissart--Gribov representation, 
$a_2^{(2)}=(2.72\pm0.36)\times10^{-4}\,M_{\pi}^{-5}.
$
We  get 
$$B_0=(2.4\pm0.3)\times10^3,\quad B_1=(6.8\pm0.8)\times10^3,\quad
 B_2=(23.7\pm3.8)\times10^3,\quad
\deltav=196\pm20\,\mev.
\equn{(2.4b)}$$

  The fit (\fig~3) returns  reasonable
 numbers for the scattering length and  for the effective range parameter,
$b_2^{(2)}$: 
$$a_2^{(2)}=(2.5\pm0.9)\times10^{-4}\,{M_\pi}^{-5};\quad
b_2^{(2)}=(-2.7\pm0.8)\times10^{-4}\,{M_\pi}^{-7}.
\equn{(2.5)}$$

\booksection{3. The high energy ($s^{1/2}\geq 1.42\,\gev$) input}

\noindent
In order to test the various dispersive sum rules we need an input for the 
 imaginary part of the scattering amplitude at high
 energy ($s^{1/2}\geq1.42\;\gev$) for which we here will 
 take  
the standard Regge theory, with 
the parameters of the various Regge poles as given in ref.~4 
(for the rho residue, we in fact take the improved version of ref.~5, Appendix~B).
We will not repeat here this Regge 
description  but will note that, in the early 1970s, Pennington and Protopopescu 
(see e.g. ref.~13)  
used crossing sum rules 
and then-existing low energy phase shifts data to shed doubt on the 
standard Regge picture for $\pi\pi$ scattering.\fnote{It is appropriate to 
remark that such doubts could  perhaps be maintained in 1974, 
when the $\pi\pi$ phase shift were  poorly known and, 
above all, it was 
also not clear which was the correct theory of strong interactions. 
They are of course difficult to believe once it became
 clear that the standard Regge picture is a feature of QCD.}
This was then accepted unquestioningly by ACGL and CGL, who 
adopted a very distorted version of Regge behaviour with, for example, 
 a Pomeron a {\sl third} of what factorization 
(or experimental data on the total $\pi\pi$ cross section) implies, 
and unconventional slopes as well.  This was  
 used by ACGL, CGL in their analyses. 
As discussed in  refs.~4,~5, however, standard Regge
 theory is perfectly consistent with
crossing  sum rules, as well as with  experiment\ref{14} if assumed to hold
systematically above $1.42\;\gev$.

\midinsert{
\medskip
\centerline{
\epsfxsize 7.8truecm\epsfbox{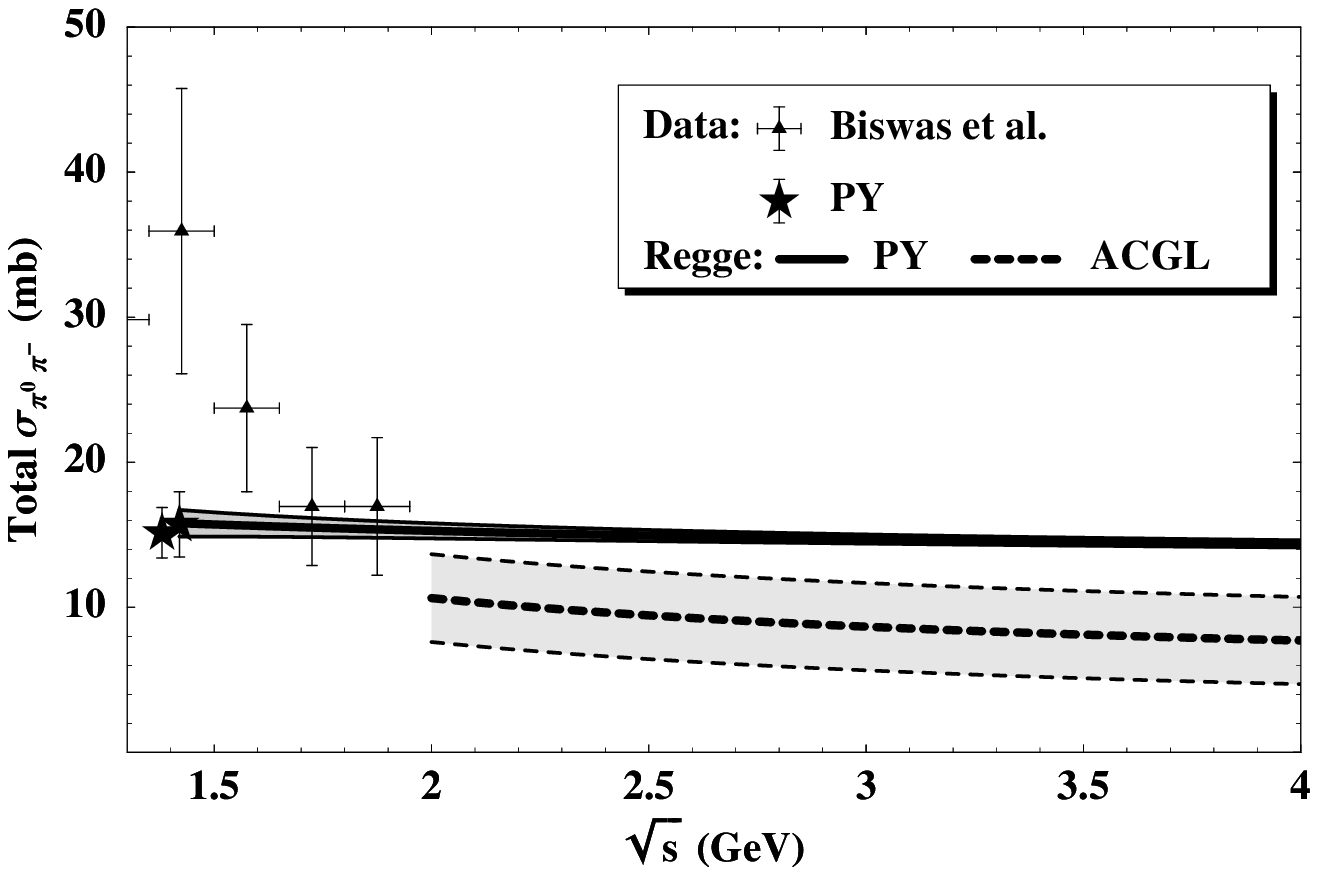}
\epsfxsize 7.8truecm\epsfbox{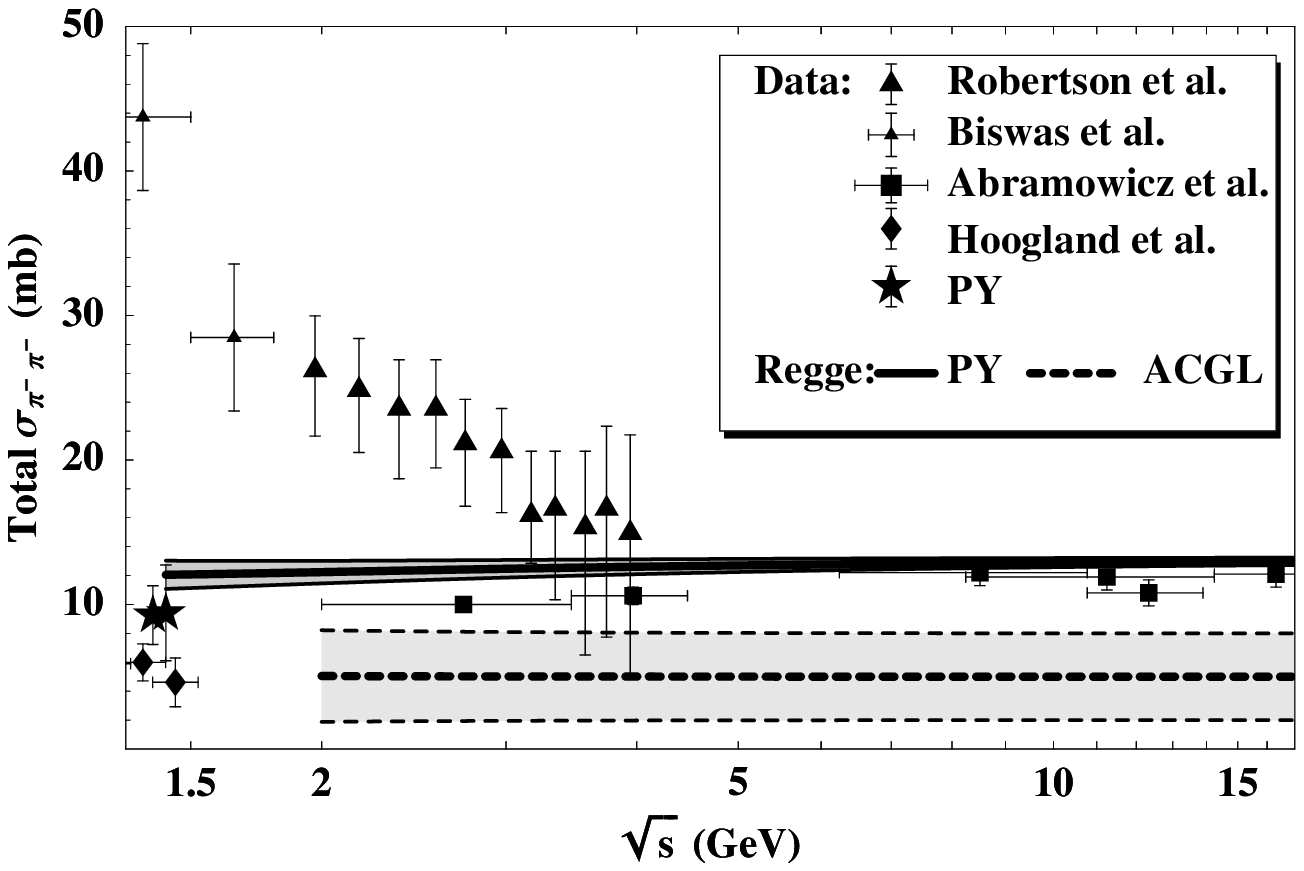}}
\medskip
\centerline{
\epsfxsize 13truecm\epsfbox{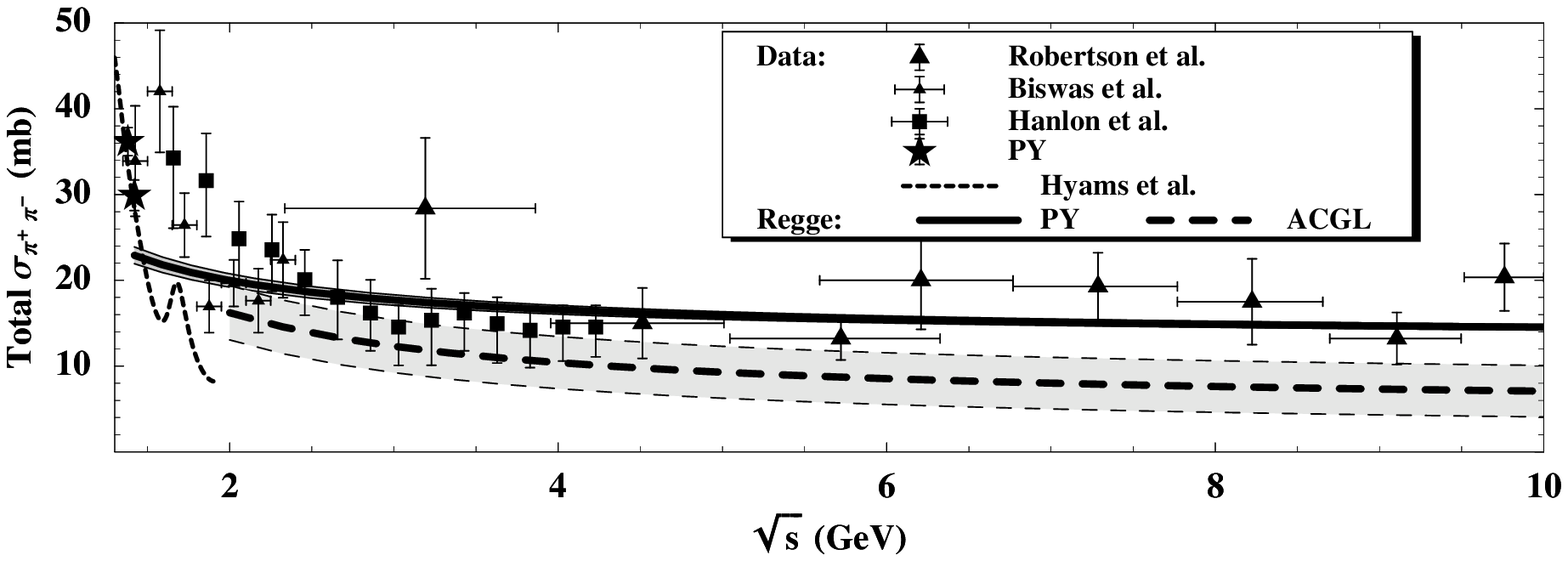}}
\bigskip
\setbox5=\vbox{\hsize 14truecm\captiontype\figurasc{Figure 4. }
 {The $\pi\pi$ cross sections. Experimental points from ref.~14. 
The stars at 1.38 and 1.42 \gev\ (PY) are from the phase shift analysis 
of experimental data given in ref.~5. 
Continuous lines, from 1.42 \gev\ (PY): Regge formula, with parameters as in 
refs.~4,~5
(the three lines per fit  cover the error in the theoretical values of the 
Regge residues). 
Dashed lines, above 2 \gev: the cross sections following from ACGL;\ref{1} 
the grey band covers their error band.
Below 2 \gev, the dotted line 
corresponds to  the $\pi^+\pi^-$ cross section from 
 the Cern--Munich analysis; cf.~\fig~7 in the paper of Hyams~et~al.\ref{10a}}}
\centerline{\box5}
\bigskip
}\endinsert

In \fig~4 we show the Regge values for the cross sections of
 the various $\pi\pi$ scattering
amplitudes, as given in 
refs.~4,~5, together with experimental data,\ref{14}
 and what follows from the imaginary parts used by ACGL\ref{1} and
CGL.\ref{2} 
It is not necessary to comment on this figure, 
which speaks for itself.

We next remark that
ACGL and CGL do {\sl not} use Regge theory for $1.42\,\gev\lsim s^{1/2}\leq2\,\gev$, 
but take here the scattering amplitude as reconstructed from 
the phase shift analyses of the Cern--Munich group.\fnote{For the S0 wave ACGL take the  
re-elaboration of the Cern--Munich data by Au, Morgan and Pennington.\ref{10d}} 
Unfortunately,  
in this region inelasticty is large and the Cern--Munich experiments, which only measure the 
{\sl differential} cross section for 
$\pi\pi\to\pi\pi$, are insufficient to reconstruct without ambiguity 
the full imaginary part. This is discussed in some detail
 in ref.~3 and, especially, in Appendix~C of ref.~5 
where it is shown that the Cern--Munich phases fail to pass a number of consistency tests. 
This is also seen very clearly in \fig~4, where we plot the total 
cross section for $\pi^+\pi^-$ that follows from the analysis of Hyams et al.,\ref{10a}
which ACGL and CGL follow. Between $s^{1/2}\sim1.5$ and $s^{1/2}\sim2\,\gev$, it 
 is clearly incompatible with the other experimental data,\ref{14} 
as well as with Regge behaviour (as follows
from factorization or fit to $\pi\pi$ data).

The situation is even worse for some specific waves. 
In particular, we will say a few words on the D2 wave.
 For D2, ACGL and CGL take (Appendix~B.3 in ACGL) an empirical fit:
$$\delta_2^{(2)}(s)=-0.003(s/4M^2_\pi)\left(1-4M^2_\pi/s\right)^{5/2}.
\equn{(3.1)}$$

Eq.~(3.1) was borrowed by ACGL from an old fit in the book of 
Martin, Morgan and Shaw,\ref{15} where only intermediate energy data were fitted. 
In fact, (3.1) fails at threshold (it gives a negative scattering length) 
and it does not fit well  experimental data below 1.42~\gev, as 
shown in \fig~3. 
Above 1~\gev, the D2 wave in (3.1) grows (in modulus) 
quadratically with the energy, while from Regge theory it is easy to verify 
that all waves should go to a multiple of $\pi$ 
and, in particular, D2 should go to zero; see
Appendix~C to ref.~5 for details. It is true that this D2 wave is small but, given the accuracy
claimed by CGL,  it is certainly not negligible.

It follows that the imaginary parts of the scattering amplitudes that ACGL, CGL use above 
1.42~\gev\ are noticeably different from the physical ones. 
We will later see how much this affects their output.

\booksection{4. Improving the  parametrizations with the help of dispersion 
relations}
\vskip-0.5truecm
\booksubsection{4.1. Checking forward dispersion relations}

\noindent 
We expect that the scattering amplitudes that follow from the experimental 
phase shifts (and inelasticities) at low energy ($s^{1/2}\leq1.42\,\gev$), 
together with the Regge expressions 
at high energy, will provide $\pi\pi$ amplitudes that
 satisfy dispersion relations since they 
fit well the experimental data and are therefore good approximations to the 
physical scattering amplitudes. 
In the present Section we will check that this is the case, 
at low energies ($s^{1/2}\lsim1\,\gev$), for 
three independent scattering amplitudes, which we 
will conveniently take the following $t$-symmetric or antisymmetric combinations, 
that form a complete set: 
$\pi^0\pi^0\to\pi^0\pi^0$, $\pi^0\pi^+\to\pi^0\pi^+$, and the 
amplitude $I_t=1$,  corresponding to isospin unity in the $t$ channel.
The reason for choosing these amplitudes is that 
the amplitudes for $\pi^0\pi^0$ and $\pi^0\pi^+$ 
depend  only on two isospin states, and have positivity properties: 
their imaginary parts are sums of positive terms. 
Because of this, the {\sl errors} are much reduced for them. 

We therefore consider here the following dispersion relations: for $\pi^0\pi^0$,
$$\real F_{00}(s)-F_{00}(4M_{\pi}^2)=
\dfrac{s(s-4M_{\pi}^2)}{\pi}\pepe\int_{4M_{\pi}^2}^\infty\dd s'\,
\dfrac{(2s'-4M^2_\pi)\imag F_{00}(s')}{s'(s'-s)(s'-4M_{\pi}^2)(s'+s-4M_{\pi}^2)}.
\equn{(4.1a)}$$
In particular, for $s=2M^2_\pi$, which will be important when we later discuss 
the Adler zeros (\subsect~4.2), we have
$$F_{00}(4M_{\pi}^2)=F_{00}(2M_{\pi}^2)+D_{00},\quad D_{00}=
\dfrac{8M_{\pi}^4}{\pi}\int_{4M_{\pi}^2}^\infty\dd s\,
\dfrac{\imag F_{00}(s)}{s(s-2M_{\pi}^2)(s-4M_{\pi}^2)}.
\equn{(4.1b)}$$
For $\pi^0\pi^+$:
$$\real F_{0+}(s)-F_{0+}(4M_{\pi}^2)=
\dfrac{s(s-4M^2_\pi)}{\pi}\pepe\int_{4M_{\pi}^2}^\infty\dd s'\,
\dfrac{(2s'-4M^2_\pi)\imag F_{0+}(s')}{s'(s'-s)(s'-4M_{\pi}^2)(s'+s-4M_{\pi}^2)}.
\equn{(4.2a)}$$
At the point $s=2M^2_\pi$, this becomes
$$F_{0+}(4M_{\pi}^2)=F_{0+}(2M_{\pi}^2)+D_{0+},\quad
D_{0+}=\dfrac{8M_{\pi}^4}{\pi}
\int_{8M_{\pi}^2}^\infty\dd s\,\dfrac{\imag F_{0+}(s)}{s(s-2M_{\pi}^2)(s-4M_{\pi}^2)}.
\equn{(4.2b)}$$
Finally, for isospin unit exchange, which does not require subtractions, 
$$\real F^{(I_t=1)}(s,0)=\dfrac{2s-4M^2_\pi}{\pi}\,\pepe\int_{4M^2_\pi}^\infty\dd s'\,
\dfrac{\imag F^{(I_t=1)}(s',0)}{(s'-s)(s'+s-4M^2_\pi)}. 
\equn{(4.3)}$$
The imaginary parts at high energy, $s^{1/2}\geq 1.42\,\gev$,
 are taken from Regge theory as discussed in refs.~4,~5 and in 
the previous Section.
Depending on the experimental data we use to fit the S0 wave we find the results reported 
 in Table~1 for the fulfillment of the dispersion 
relations. The \chidof\ is obtained 
by evaluating left and right hand sides of Eqs.~(4.1a), (4.2a) and (4.3) 
at intervals of 25~\mev, from threshold to 0.925~\gev, 
adding the $\chi^2$'s for all points, and dividing by the number of  points. 
We do not give, in Table~1, the \chidof\ corresponding 
to $\pi^0\pi^+$ scattering, which is equal to 1.7 independently of the S0 wave.

We also give in Table~1 the central values of the quantity $\delta_0^{(0)}(0.8^2\,\gev^2)$, that
we  will discuss later.

\midinsert
{\medskip
\setbox0=\vbox{
\setbox1=\vbox{\petit \offinterlineskip\hrule
\halign{
&\vrule#&\strut\hfil\ #\ \hfil&\vrule#&\strut\hfil\ #\ \hfil&
\vrule#&\strut\hfil\ #\ \hfil&
\vrule#&\strut\hfil\ #\ \hfil&
\vrule#&\strut\hfil\ #\ \hfil&
#&\strut\hfil\ #\ \hfil&
\vrule#&\strut\hfil\ #\ \hfil&\vrule#&\strut\hfil\ #\ \hfil\cr
 height2mm&\omit&&\omit&&\omit&&\omit&&\omit&&\omit&&\omit&&\omit&\cr 
&\hfil \hfil&&\hfil $B_0$ \hfil&&\hfil $B_1$\hfil&&\hfil $M_\sigma$ (MeV)\hfil&&
&&\hfil $\dfrac{\chi^2}{\rm d.o.f.}(I_t=1)$\hfil&&\hfil 
 $\dfrac{\chi^2}{\rm d.o.f.}(\pi^0\pi^0)$\hfil&&
$\delta_0^{(0)}(0.8^2)$& \cr
 height1mm&\omit&&\omit&&\omit&&\omit&&\omit&&\omit&&\omit&&\omit&\cr
\noalign{\hrule} 
height1mm&\omit&&\omit&&\omit&&\omit&&\omit&&\omit&&\omit&&\omit&\cr
&PY, Eq. (2.14)&&\vphantom{\Big|}$21.04$&&
$6.62$&&$782\pm24$&&& &\hfil$0.3$ \hfil&&
$3.5$&&
$91.9\degrees$& \cr 
\noalign{\hrule}
height1mm&\omit&&\omit&&\omit&&\omit&&\omit&&\omit&&\omit&&\omit&\cr
&\vphantom{\Big|}$K\; {\rm decay\;  only}$
&&$\phantom{\Big|}18.5\pm1.7$&&$\equiv0$&&$766\pm95$&&&
&\hfil $0.2$  \hfil&&$1.8$&&
$93.2\degrees$& \cr
\noalign{\hrule}
height1mm&\omit&&\omit&&\omit&&\omit&&\omit&&\omit&&\omit&&\omit&\cr
&\vphantom{\Big|}
${\displaystyle{{ K\; {\rm decay\; data}}}\atop{\displaystyle +\,{\rm Grayer,\;B}}}$&&
$22.7\pm1.6$ &&$12.3\pm3.7 $&&$858\pm15 $&
&&
&$1.0$&&$2.7$&&
$84.0\degrees$& \cr
\noalign{\hrule}
height1mm&\omit&&\omit&&\omit&&\omit&&\omit&&\omit&&\omit&&\omit&\cr
&\vphantom{\Big|}
${\displaystyle{{ K\; {\rm decay\; data}}}\atop{\displaystyle +\,{\rm Grayer,\;C}}}$&&
$ 16.8\pm0.85$&&$-0.34\pm2.34 $&&$787\pm9 $&
&&
&$0.4$&&$1.0$&&
$91.1\degrees$& \cr
\noalign{\hrule}
height1mm&\omit&&\omit&&\omit&&\omit&&\omit&&\omit&&\omit&&\omit&\cr
&\vphantom{\Big|}
${\displaystyle{{ K\; {\rm decay\; data}}}\atop{\displaystyle +\,{\rm Grayer,\;E}}}$&&
$21.5\pm3.6 $&&$12.5\pm7.6 $&&$1084\pm110 $&
&&
&$2.1$&&$0.5$&&
$70.6\degrees$& \cr
\noalign{\hrule}
height1mm&\omit&&\omit&&\omit&&\omit&&\omit&&\omit&&\omit&&\omit&\cr
&\vphantom{\Big|}
${\displaystyle{{ K\; {\rm decay\; data}}}\atop{\displaystyle +\,{\rm Kaminski}}}$&&
$ 27.5\pm3.0$&&$21.5\pm7.4 $&&$789\pm18 $&
&&
&$0.3$&&$5.0$&&
$91.6\degrees$& \cr
\noalign{\hrule}
height1mm&\omit&&\omit&&\omit&&\omit&&\omit&&\omit&&\omit&&\omit&\cr
\noalign{\hrule}
height1mm&\omit&&\omit&&\omit&&\omit&&\omit&&\omit&&\omit&&\omit&\cr
&\vphantom{\Big|}
${\displaystyle{{ K\; {\rm decay\; data}}}\atop{\displaystyle +\,{\rm Grayer,\;A}}}$&&
$ 28.1\pm1.1$&&$26.4\pm2.8 $&&$866\pm6 $&
&&
&$2.0$&&$7.9$&&
$81.2\degrees$& \cr
\noalign{\hrule}
height1mm&\omit&&\omit&&\omit&&\omit&&\omit&&\omit&&\omit&&\omit&\cr
&\vphantom{\Big|}
${\displaystyle{{ K\; {\rm decay\; data}}}\atop{\displaystyle +\,{{\rm EM},\;s{\rm -channel}}}}$&&
$ 29.8\pm1.3$&&$25.1\pm3.3 $&&$811\pm7 $&
&&
&$1.0$&&$9.1$&&
$88.3\degrees$& \cr
\noalign{\hrule}
height1mm&\omit&&\omit&&\omit&&\omit&&\omit&&\omit&&\omit&&\omit&\cr
&\vphantom{\Big|}
${\displaystyle{{ K\; {\rm decay\; data}}}\atop{\displaystyle +\,{{\rm EM},\;t{\rm -channel}}}}$&&
$ 29.3\pm1.4$&&$26.9\pm3.4 $&&$829\pm6 $&
&&
&$1.2$&&$10.1$&&
$85.7\degrees$& \cr
\noalign{\hrule}
height1mm&\omit&&\omit&&\omit&&\omit&&\omit&&\omit&&\omit&&\omit&\cr
&\vphantom{\Big|}
${\displaystyle{{ K\; {\rm decay\; data}}}\atop{\displaystyle +\,{\rm Protopopescu\,VI}}}$&&
$ 27.0\pm1.7$&&$22.0\pm4.1 $&&$855\pm10 $&
&&
&$1.2$&&$5.8$&&
$82.9\degrees$& \cr
\noalign{\hrule}
height1mm&\omit&&\omit&&\omit&&\omit&&\omit&&\omit&&\omit&&\omit&\cr
&\vphantom{\Big|}
${\displaystyle{{ K\; {\rm decay\; data}}}\atop{\displaystyle +\,{\rm Protopopescu\,XII}}}$&&
$ 25.5\pm1.7$&&$18.5\pm4.1 $&&$866\pm14 $&
&&
&$1.2$&&$6.3$&&
$82.2\degrees$& \cr
\noalign{\hrule}
height1mm&\omit&&\omit&&\omit&&\omit&&\omit&&\omit&&\omit&&\omit&\cr
&\vphantom{\Big|}
${\displaystyle{{ K\; {\rm decay\; data}}}\atop{\displaystyle +\,{\rm Protopopescu\,3}}}$&&
$ 27.1\pm2.3$&&$23.8\pm5.0 $&&$913\pm18 $&
&&
&$1.8$&&$4.2$&&
$76.7\degrees$& \cr
\noalign{\hrule}}
\vskip.05cm}
\centerline{\box1}
\bigskip
{\noindent\petit
 PY, Eq.~(2.3): our 
global fit, \equn{(2.3)}.  We do not give errors here as they are strongly correlated.
Grayer A, B, C, E: the corresponding solutions in the 
paper of Grayer et al.\ref{10b} (note that solution D in this paper concerns only 
data at higher energies). EM: the
solutions of Estabrooks and  Martin.\ref{10e}  Kaminski: the papers of 
Kamin\'ski et al.\ref{10f}  Protopopescu VI, XII and VIII:  
 the corresponding solutions in ref.~9.}
\medskip
\centerline{\sc Table~1}
\smallskip
\centerrule{5truecm}}
\box0
}
\endinsert

\booksubsection{4.2. Improving the fits using forward dispersion relations}

\noindent
In this Section we will improve the values of the 
parameters we have found with our fits to data requiring also 
 fulfillment (within errors) of 
 dispersion relations up to 0.925~\gev. 
This will provide us with a parametrization of the various waves with central values 
more compatible with analyticity and $s\,-\,u$ crossing. 
This method is an alternative to that of the Roy equations to which it is
 superior in that we do not need as input the 
values of the scattering amplitude for
$|t|$ up to
$30M^2_\pi$, where the various Regge fits  existing in the literature disagree strongly
one with another (see ref.~5, Appendix~B),
 and also in that, with  dispersion relations, we can 
test all energies, whereas the Roy equations are only  valid for
$s^{1/2}<\sqrt{60}\,M_\pi\sim 1.1\,\gev$
(and in practice only applied up to $0.8\,\gev$). 
With the S0 
wave in \equn{(2.3)}, we find (units of $M_\pi$)
$$\eqalign{
{\rm S0};\; s^{1/2}\leq 2m_K:&\quad  B_0=17.4\pm0.5;\quad B_1=4.3\pm1.4;
\cr &\quad M_\sigma=790\pm21\,\mev;\quad
z_0=195\,\mev\;\hbox{[Fixed]};\cr
&\quad  a_0^{(0)}=0.230\pm0.015;\quad b_0^{(0)}=0.312\pm0.014.\cr
{\rm S2};\; s^{1/2}\leq 1.0\,\gev:&\quad B_0=-80.8\pm1.7;\quad B_1=-77\pm5;\quad
z_2=147\,\mev\;\hbox{[Fixed]};\cr
&\quad  a_0^{(2)}=-0.0480\pm0.0046;\quad b_0^{(2)}=-0.090\pm0.006.\cr
{\rm S2};\;  1.0\leq s^{1/2}\leq1.42:&\quad  B_0=-125\pm6;\quad B_1=-119\pm14;\quad
\epsilon=0.17\pm0.12.\cr
{\rm P};\; s^{1/2}\leq 1.05:&\quad 
B_0=1.064\pm0.11;\quad B_1=0.170\pm0.040;\quad
M_\rho=773.6\pm0.9\;\mev;\cr
&\quad a_1=(38.7\pm1.0)\times10^{-3};\quad b_1=(4.55\pm0.21)\times10^{-3}.\cr
 {\rm D0};\; s^{1/2}\leq
1.42:&\quad B_0=23.5\pm0.7;\quad B_1=24.8\pm1.0;\quad \epsilon=0.262\pm0.030;\cr
& \quad a_2^{(0)}=(18.4\pm3.0)\times10^{-4};\quad b_2^{(0)}=(-8.6\pm3.4)\times10^{-4}.
\cr  {\rm D2};\; s^{1/2}\leq 1.42:&\quad  
B_0=(2.9\pm0.2)\times10^3;\quad B_1=(6.3\pm0.8)\times10^3;\quad
B_2=(25.4\pm3.6)\times10^3;\cr
&\quad\deltav=212\pm19;\cr
&\quad a_2^{(2)}=(2.4\pm0.7)\times10^{-4};\quad b_2^{(2)}=(-2.5\pm0.6)\times10^{-4}.\cr
{\rm F};\; s^{1/2}\leq 1.42:&\quad B_0=(1.09\pm0.03)\times10^5;\quad
B_1=(1.41\pm0.04)\times10^5;
\quad a_3=(6.0\pm0.8)\times10^{-5}.\cr}
\equn{(4.4)}$$ 
When using the fits to individual sets of data we get somewhat different results, 
for the S0 wave parameters (the other waves are essentially like in (4.4), however):

\midinsert{
\setbox0=\vbox{
\setbox1=\vbox{\petit \offinterlineskip\hrule
\halign{
&\vrule#&\strut\hfil\ #\ \hfil&\vrule#&\strut\hfil\ #\ \hfil&
\vrule#&\strut\hfil\ #\ \hfil&
\vrule#&\strut\hfil\ #\ \hfil&
\vrule#&\strut\hfil\ #\ \hfil&
\vrule#&\strut\hfil\ #\ \hfil&
\vrule#&\strut\hfil\ #\ \hfil&
\vrule#&\strut\hfil\ #\ \hfil&
\vrule#&\strut\hfil\ #\ \hfil&\vrule#&\strut\hfil\ #\ \hfil\cr
 height2mm&\omit&&\omit&&\omit&&\omit&&\omit&&\omit&&\omit&&\omit&&\omit&&\omit&\cr 
&\hfil Improved fits: \hfil&&\hfil $B_0$ \hfil&&\hfil $B_1$\hfil&&\hfil $M_\sigma$ (\mev)\hfil&&
\hfil $z_0$ (\mev)\hfil&&\hfil $\dfrac{\chi^2(I_t=1)}{\rm d.o.f.}$\hfil&&
\hfil $\dfrac{\chi^2(00)}{\rm d.o.f.}$\hfil&&\hfil 
 $\dfrac{\chi^2(0+)}{\rm d.o.f.}$\hfil&&
 (4.1b)\hfil&&
 $\delta_0^{(0)}(.8^2)$\hfil& \cr
 height1mm&\omit&&\omit&&\omit&&\omit&&\omit&&\omit&&\omit&&\omit&&\omit&&\omit&\cr
\noalign{\hrule} 
height1mm&\omit&&\omit&&\omit&&\omit&&\omit&&\omit&&\omit&&\omit&&\omit&&\omit&\cr
&PY, Eq. (4.4)&&\vphantom{\Big|}$ 17.4\pm0.5$&&$4.3\pm1.4$&&$790\pm30$&&$198\pm21$&&$0.40$&
&\hfil$0.66$ \hfil&&
$1.62$&&
 $1.6\,\sigma$&&
 $91.3\degrees$\hfil& \cr 
\noalign{\hrule}
height1mm&\omit&&\omit&&\omit&&\omit&&\omit&&\omit&&\omit&&\omit&&\omit&&\omit&\cr
&\vphantom{\Big|}$K\; {\rm decay\;  only}$
&&$\phantom{\Big|}16.4\pm0.9$&&$\equiv0$&&$809\pm53$&&$182\pm34$&&$0.30$&
&\hfil $0.29$  \hfil&&$1.77$&&
$1.5\,\sigma$\hfil&&
 $91.3\degrees$\hfil& \cr
\noalign{\hrule}
height1mm&\omit&&\omit&&\omit&&\omit&&\omit&&\omit&&\omit&&\omit&&\omit&&\omit&\cr
&\vphantom{\Big|}
${\displaystyle{{ K\; {\rm decay\; data}}}\atop{\displaystyle +\,{\rm Grayer,\;C}}}$&&
$16.2\pm0.7$ &&$0.5\pm1.8 $&&$788\pm9 $&&$184\pm39 $&
&\hfil$0.37$ \hfil&
&$0.32$&&$1.74$&&
 $1.5\,\sigma$\hfil&&
 $91.0\degrees$\hfil& \cr
\noalign{\hrule}height1mm&\omit&&\omit&&\omit&&\omit&&\omit&&\omit&&\omit&&\omit&&\omit&&\omit&\cr
\noalign{\hrule}
height1mm&\omit&&\omit&&\omit&&\omit&&\omit&&\omit&&\omit&&\omit&\cr
&\vphantom{\Big|}
${\displaystyle{{ K\; {\rm decay\; data}}}\atop{\displaystyle +\,{\rm Grayer,\;B}}}$&&
$ 20.7\pm1.0$&&$11.6\pm2.6 $&&$861\pm14 $&&$233\pm30 $&
&\hfil $0.37 $ \hfil&
&$0.83$&&$1.60$&&
$4.0\,\sigma$\hfil&&
 $85.1\degrees$\hfil& \cr
\noalign{\hrule}
height1mm&\omit&&\omit&&\omit&&\omit&&\omit&&\omit&&\omit&&\omit&&\omit&&\omit&\cr
&\vphantom{\Big|}
${\displaystyle{{ K\; {\rm decay\; data}}}\atop{\displaystyle +\,{\rm Grayer,\;E}}}$&&
$ 20.2\pm2.2$&&$8.4\pm5.2 $&&$982\pm95 $&&$272\pm50 $&
&\hfil $0.60 $ \hfil&
&$0.09$&&$1.40$&&
$6.0\,\sigma$\hfil&&
 $78.0\degrees$\hfil& \cr
\noalign{\hrule}
height1mm&\omit&&\omit&&\omit&&\omit&&\omit&&\omit&&\omit&&\omit&&\omit&&\omit&\cr
&\vphantom{\Big|}
${\displaystyle{{ K\; {\rm decay\; data}}}\atop{\rm\displaystyle +\,Kaminski}}$&&
$ 20.8\pm1.4$&&$13.6\pm3.7 $&&$798\pm17 $&&$245\pm39 $&
&\hfil $0.43 $ \hfil&
&$1.08$&&$1.36$&&
$4.5\,\sigma$\hfil&&
 $91.8\degrees$\hfil& \cr
\noalign{\hrule}}
\vskip.05cm}
\centerline{\box1}
\bigskip
{\noindent\petit
 PY,~Eq.~(4.4): from the fit (2.3),
 improved with forward dispersion relations.
Other conventions  as in Table~1.
Errors are given for the Adler zero, but we fix it at its  central 
value when evaluating the errors for the other parameters.
}
\medskip
\centerline{\sc Table~2}
\smallskip
\centerrule{5truecm}}
\box0
}\endinsert

\booksection{5. The ACGL, CGL phases at the matching point, $s^{1/2}=0.8\,\gev$}

\noindent
A very important role is played, in the analyses of ACGL, CGL, by the input 
phases for the S0, S2 and P waves at the point, $s^{1/2}=0.8\,\gev$, 
where they match the solutions of the Roy equations to 
the experimental amplitude at high energy. 
The errors these authors take for the S2, P waves are reasonable; that for the 
S0 wave, which is a key quantity, is not, as we will now discuss. 

The quantity $\delta_0^{(0)}((0.8\,\gev)^2)$ is in fact given
 in  Eq.~(6.3) of ACGL as
$$\delta_0^{(0)}((0.8\,\gev)^2)=82.3\pm3.4\degrees.
\equn{(5.1)}$$
This value is also used in CGL. This 
error, $\pm3.4\degrees$, 
may be contrasted with the error estimates of ref.~5, Eqs.~(2.13),  
which  vary, for 
the data above 0.8 \gev, between 6\degrees\ and 18\degrees.

Let us see if the ACGL number is reasonable. 
ACGL consider the difference $\delta_1-\delta_0^{(0)}$ at 0.8 \gev,  
in the hope that some of the errors will cancel for this combination. 
They give a value for this quantity of
$$\delta_0^{(0)}((0.8\,\gev)^2)-\delta_1((0.8\,\gev)^2)=26.6\pm2.8\degrees.$$
Now, the Cern--Munich\ref{10} experimental values for this which ACGL quote 
(there are others) are
$$\delta_1((0.8\,\gev)^2)-\delta_0^{(0)}((0.8\,\gev)^2)=\cases{
23.4\pm4.0\degrees\quad\hbox{[Hyams et al.]} \cr
24.8\pm3.8\degrees\quad\hbox{[Estabrooks and Martin, $s$-channel]} \cr
30.3\pm3.4\degrees\quad\hbox{[Estabrooks and Martin, $t$-channel]}. \cr
}
$$

\midinsert{
\setbox0=\vbox{{\psfig{figure=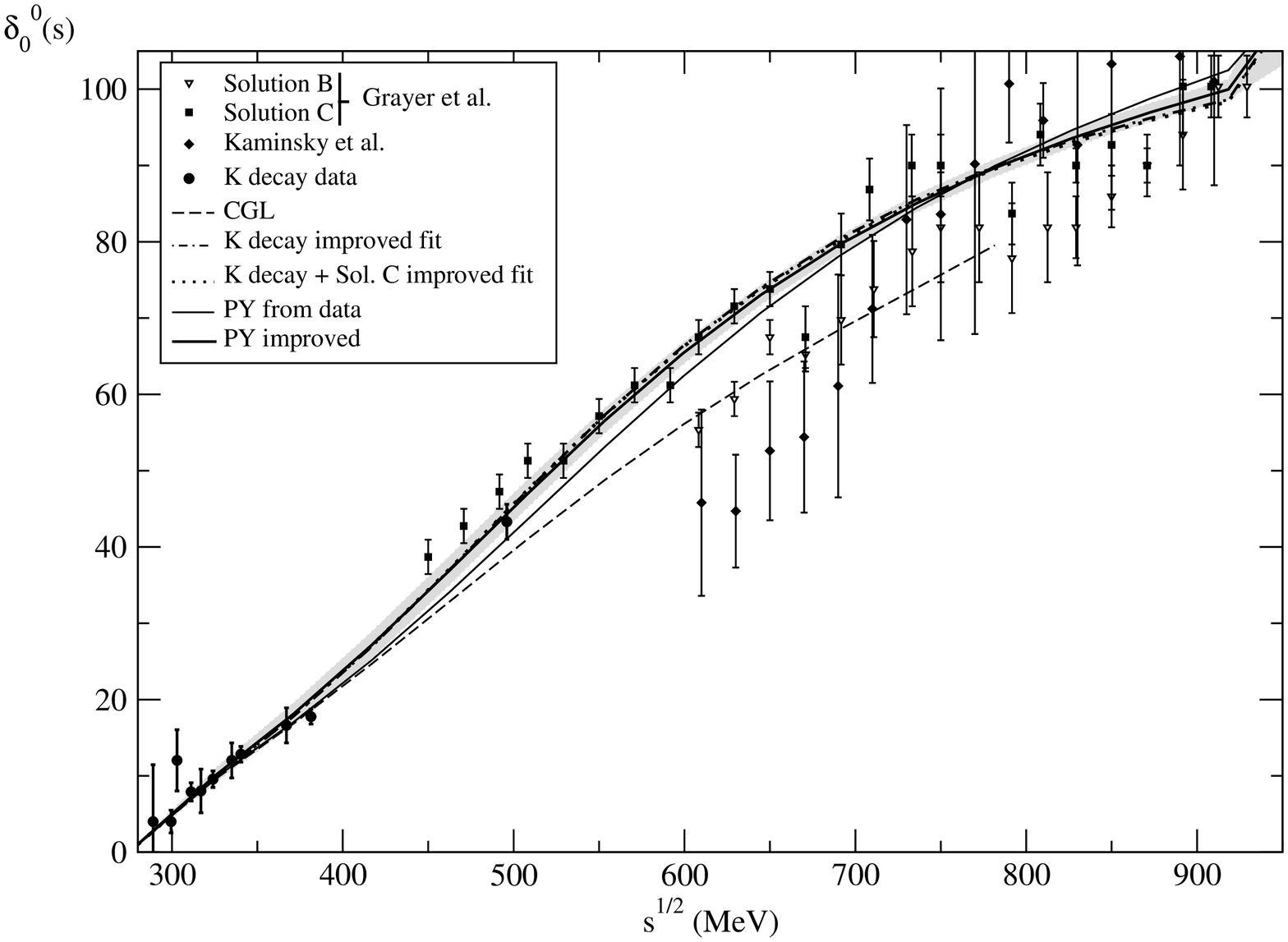,width=12.4truecm,angle=-90}}} 
\setbox6=\vbox{\hsize 14truecm\captiontype\figurasc{Figure 5. }{
The S0 phase shift 
 corresponding to the improved fit in Eq.~(4.4) (PY, thick continuous
line  
and error band), the unimproved solution of Eq.~(2.3) (thin continuous line),
 and the 
{\sl improved} solutions ``$K$ decay only" and
 ``Grayer~C" of Table~2 (difficult to see as they
fall almost on top of  PY). 
The solution CGL\ref{2} (lowest discontinuous line) is also shown. }} 
\centerline{\tightboxit{\box0}}
\bigskip
\centerline{\box6}
}\endinsert
 All the numbers here  stem from the {\sl same experiment}, 
and differ only on the method of analysis. 
Their spread is an indication of the {\sl systematic}, or theoretical, error
 in the  determination of $\delta_1-\delta_0^{(0)}$. 
To take this into account a correct procedure is to average the 
various determinations and  enlarge the error, with what we will call
 ``systematic error", so that all the   
numbers  are covered by it:
$$\delta_1((0.8\,\gev)^2)-\delta_0^{(0)}((0.8\,\gev)^2)=
26.3\degrees\pm3.8\,({\rm St.})\pm4.0\,({\rm Sys.)}.$$
Using now ACGL's central value 
$\delta_1((0.8\,\gev)^2)=108.9\pm2\degrees$ 
we would find
$$\delta_0^{(0)}((0.8\,\gev)^2)=82.7\pm7.8\degrees;
\equn{(5.2)}
$$
we have added (St.) and (Sys.) errors linearly, as is mandatory when  a single datum 
is involved.
In fact, this is probably optimistic. 
First of all, the Hyams et al. value quoted above 
 is only one of {\sl five} 
solutions by the same experiment (depending on the corrections 
applied: cf.~Grayer et al.\ref{10b}). Secondly,   if we included the 
 data of Protopopescu et al.,\ref{9} 
 the error  would increase to $10\degrees$.  
 In any case,  a realistic error should be 
several times what ACGL and CGL  assume. 

But the best way to see that the error, and indeed, even the central value, given
 in (5.1) are not reasonable is to look at the {\sl experimental} 
results reported in
Tables~1,~2 for
$\delta_0^{(0)}((0.8\,\gev)^2)$, which speak for themselves. 

\topinsert{
\setbox0=\vbox{{\psfig{figure=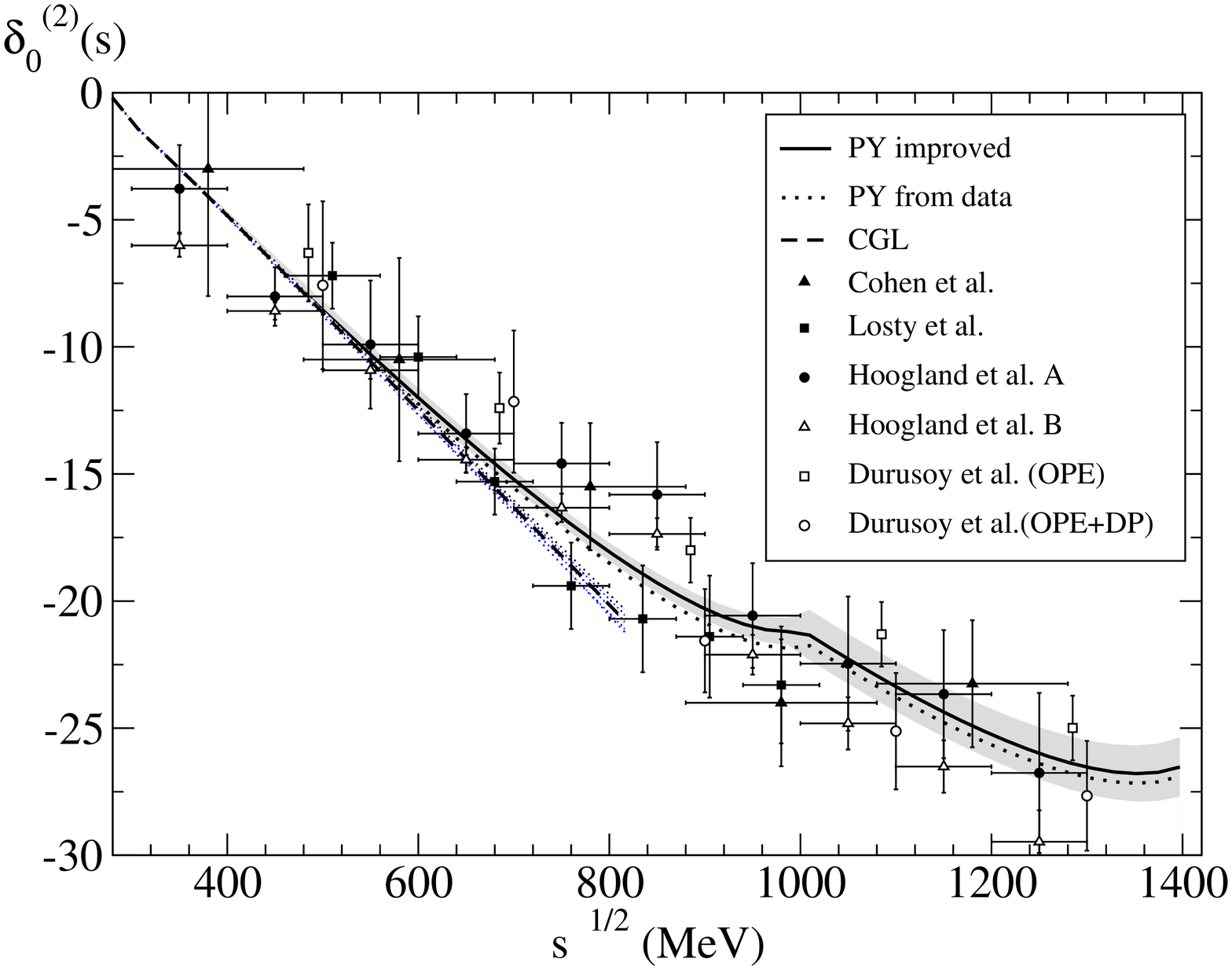,width=9.7truecm,angle=-90}}}
\setbox6=\vbox{\hsize 14truecm\captiontype\figurasc{Figure 6. }{
Phase shift for the S2 wave,  
  Eq.~(4.4) (PY, thick continuous
line, and error band); unimproved fit, Eq.~(2.2) (PY, dotted line); and 
the phase CGL,\ref{2} and  error band  (dashed line).}} 
\centerline{\tightboxit{\box0}}
\medskip
\centerline{\box6}
}\endinsert

The character of a forced fit is also indicated in Fig.~5. 
CGL  constrain the curve to 
pass through the point at (5.1). The corresponding solution 
is then dragged away from all solutions that are consistent with dispersion relations.
In fact, it probably the distortion of the S0 wave, together with the distortion  
caused by incorrect high energy behaviour, what also drives the 
 S2 wave of CGL away from what one finds from experiment (\fig~6).

\booksection{6. Dispersion relations and the CGL solution}

\noindent
The inconsistency of the CGL solution with experimental results at high energy 
is particularly transparent if we 
consider the fulfillment of dispersion
relations with the parameters of CGL for the S0, S2 and P waves al low energy, 
or with our parameters here.
\goodbreak

\topinsert{
\setbox0=\vbox{\hsize15truecm
\setbox1=\vbox{{\psfig{figure=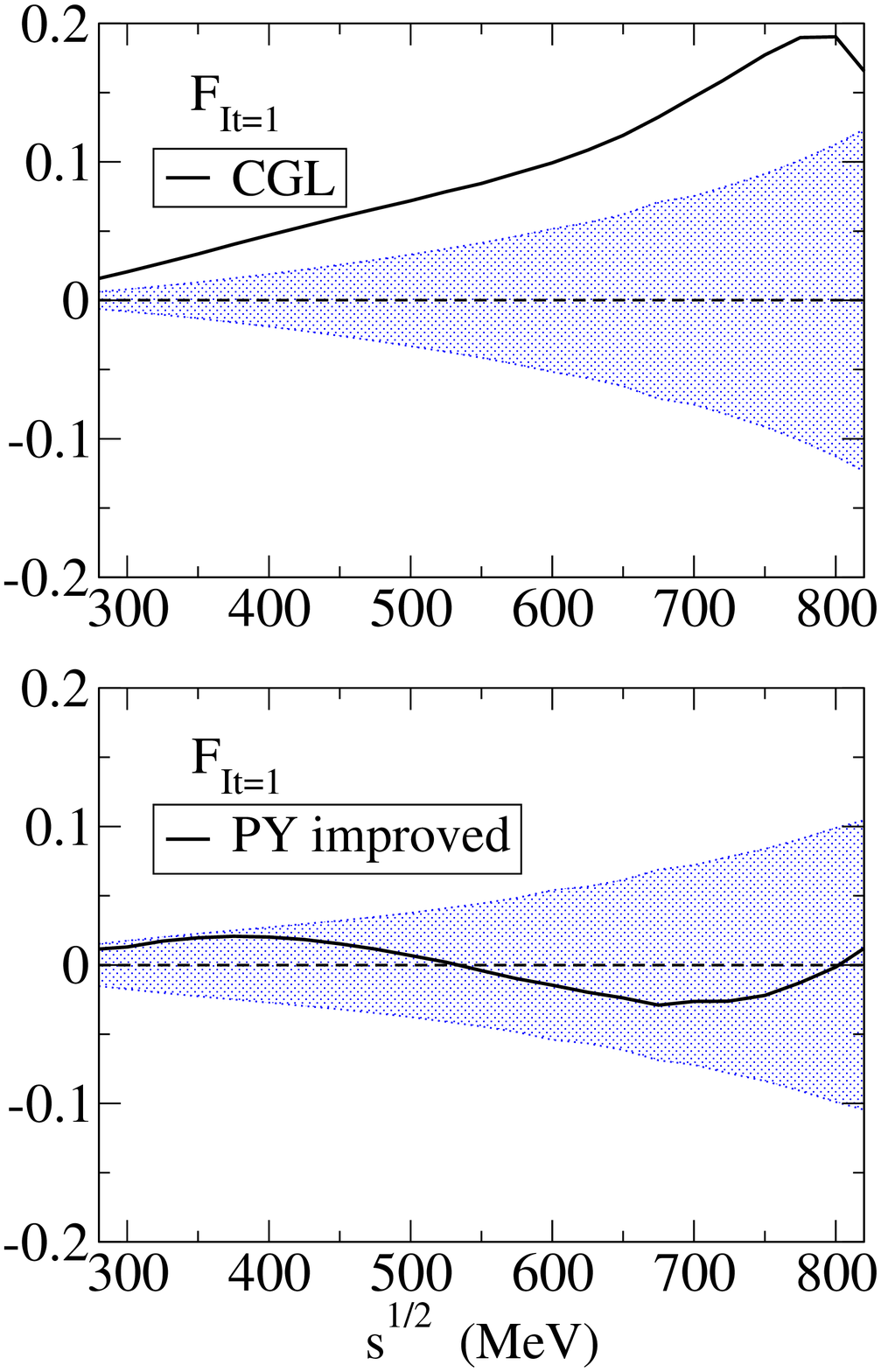,width=6.2truecm,angle=-0}}}
\setbox2=\vbox{{\psfig{figure=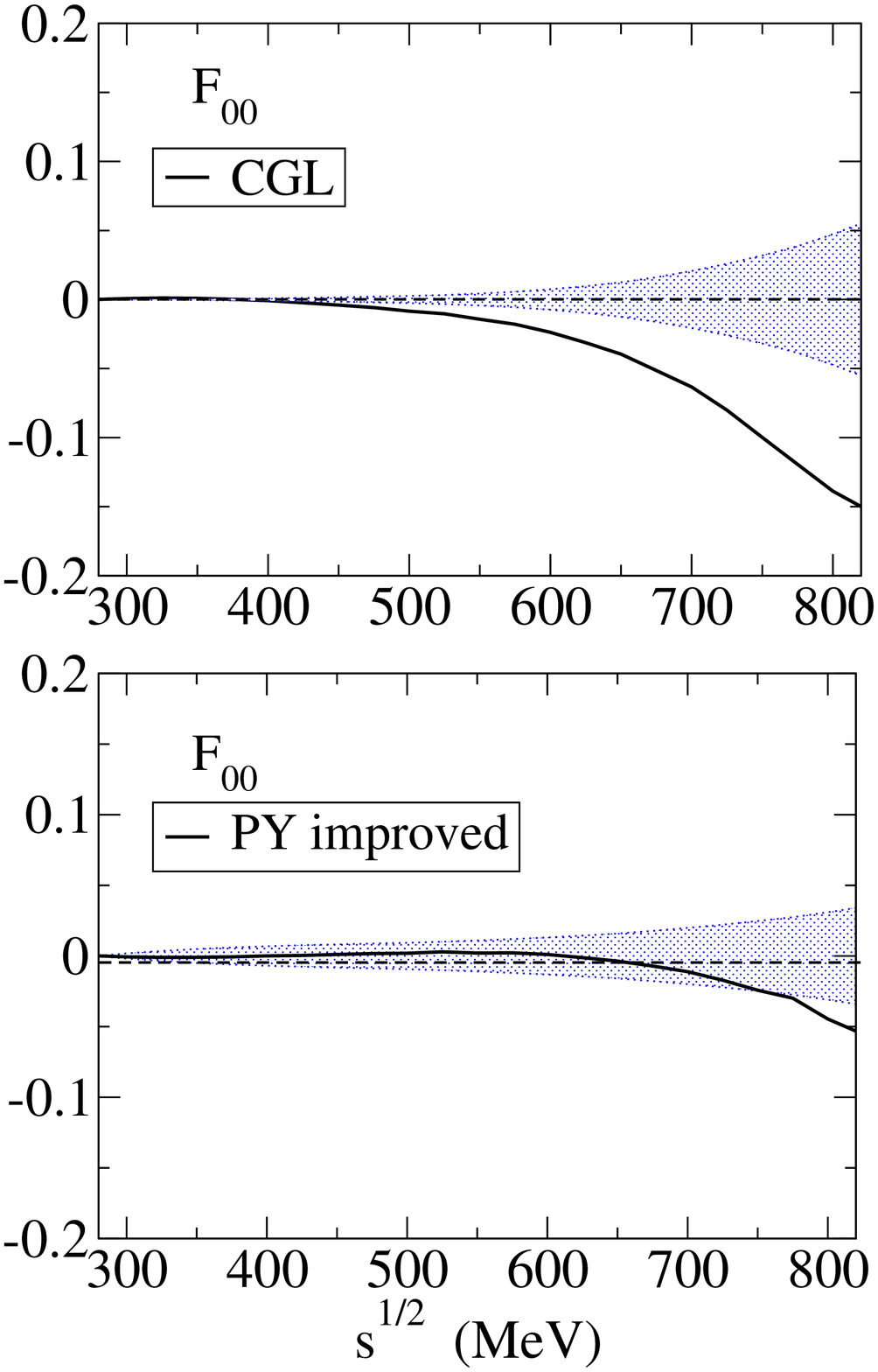,width=6.2truecm,angle=-0}}}
\setbox3=\vbox{{\psfig{figure=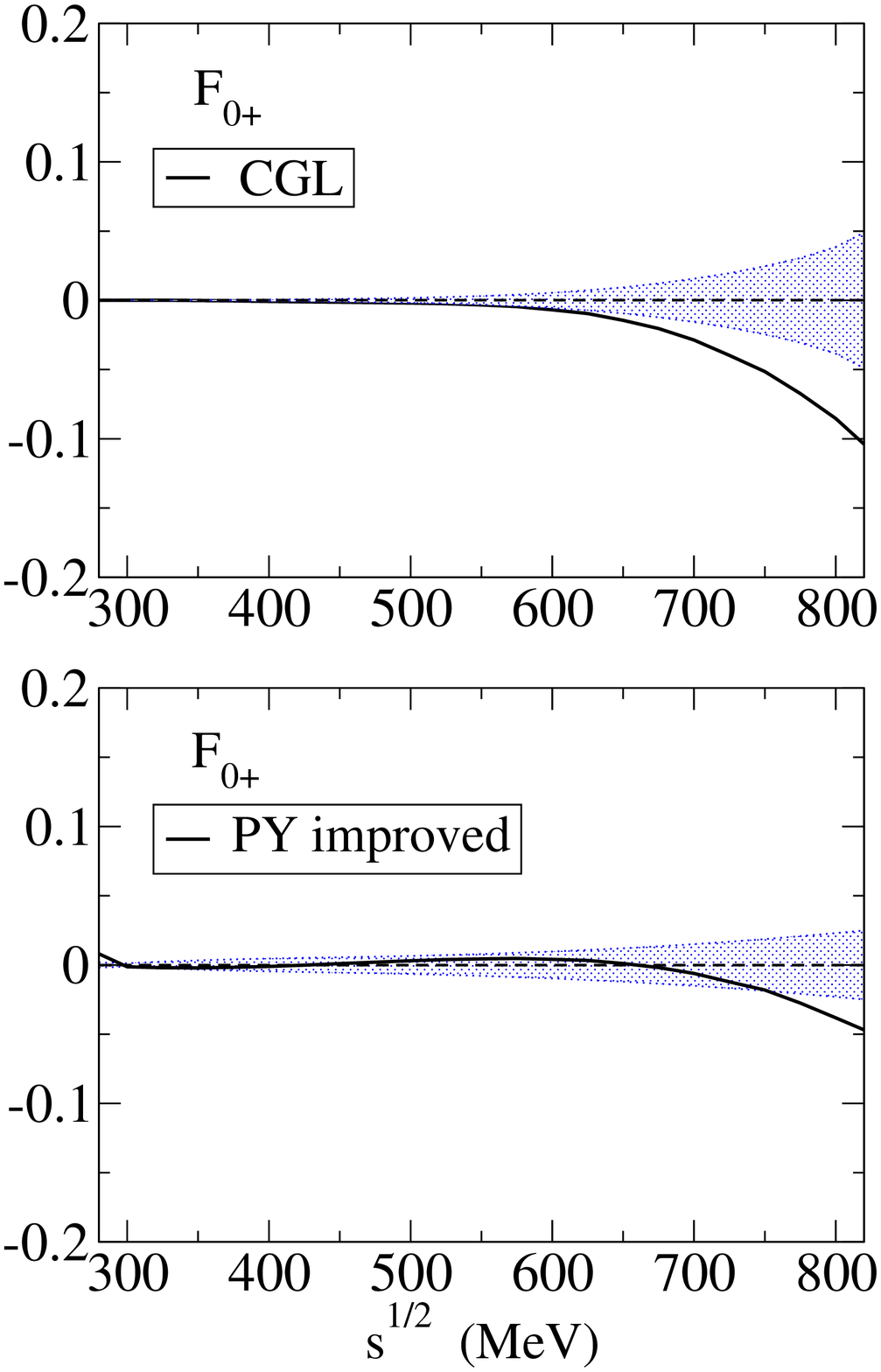,width=6.2truecm,angle=-0}}} 
\line{\tightboxit{\box1}\hfil\tightboxit{\box2}}
\setbox6=\vbox{\hsize 6truecm\captiontype\figurasc{Figure 7. }{Consistency of 
dispersion
relations for the 
$\pi\pi$ amplitudes 
of ref.~2~(CGL) and for our amplitudes, 
with the  parameters in (4.4),
denoted by PY.  We plot the differences $\deltav_i$, given in 
Eqs.~(6.1), between the results of the 
calculation of the real parts directly 
with the various parametrizations, or from the 
dispersive formulas. Perfect consistency 
would occur if the continuous curves coincided with the dotted lines.
The error bands are also shown. 
The progressive deterioration of the CGL results as the energy
 increases is apparent
here.}\hb
\vskip0.5truecm
\phantom{x}}
\line{\tightboxit{\box3}\hfil{\box6}}}
\centerline{{\box0}}
}\endinsert

This is depicted in \fig~7, where we show the mismatch between the real part and the
dispersive 
evaluations, that is to say, the differences $\deltav_i$, 
$$\deltav_{1}\equiv\real
F^{(I_t=1)}(s,0)-\dfrac{2s-4M^2_\pi}{\pi}\pepe\int_{4M^2_\pi}^\infty\dd s'\,
\dfrac{\imag F^{(I_t=1)}(s',0)}{(s'-s)(s'+s-4M^2_\pi)},
\equn{(6.1a)}$$
$$\deltav_{00}\equiv\real F_{00}(s)-F_{00}(4M_{\pi}^2)-
\dfrac{s(s-4M_{\pi}^2)}{\pi}\pepe\int_{4M_{\pi}^2}^\infty\dd s'\,
\dfrac{(2s'-4M^2_\pi)\imag F_{00}(s')}{s'(s'-s)(s'-4M_{\pi}^2)(s'+s-4M_{\pi}^2)},
\equn{(6.1b)}$$
and
$$\deltav_{0+}\equiv\real F_{0+}(s)-F_{0+}(4M_{\pi}^2)-
\dfrac{s(s-4M^2_\pi)}{\pi}\pepe\int_{4M_{\pi}^2}^\infty\dd s'\,
\dfrac{(2s'-4M^2_\pi)\imag F_{0+}(s')}{s'(s'-s)(s'-4M_{\pi}^2)(s'+s-4M_{\pi}^2)}.
\equn{(6.1c)}$$
These quantities would vanish,  $\deltav_i=0$, 
if the dispersion relations were exactly satisfied.
 
We  include in the comparison of \fig~7 the errors; in the case of CGL, 
these errors are as follow from the parametrizations given by these authors in ref.~2, 
for $s^{1/2}\lsim0.8\,\gev$. At higher energies they are taken from experiment via our 
parametrizations.
By comparison, we show the same quantities for our 
best results in the present paper, that is to say, with amplitudes  
improved by use of dispersion relations, \equn{(4.4)}. 
In both cases we have taken the Regge parameters from 
Appendix~B in ref.~5.

\brochuresection{7. Comparison of low energy parameters of CGL, DFGS, KLL, and here}

\noindent
We here present, in Table~3,  the low energy parameters as given in CGL,  as well as 
in two other recent 
evaluations, that use  the Roy equations, by Descotes et al.\ref{15}, that we denote by
DFGS, and by Kami\'nski, Le\'sniak and
Loiseau\ref{15}, denoted by KLL. 
This is compared with what one finds fitting experimental data,
 improved with dispersion relations, as reported in ref.~5 and repeated here, 
that we denote by PY. 
 The mismatches between many of the parameters of CGL and PY are apparent 
in Table~3; not surprisingly, these mismatches 
affect mostly those parameters 
which are sensitive to high energy: $b_1$, $a_2^{(I)}$,  $b_2^{(I)}$.

\midinsert{
\setbox0=\vbox{\petit
\setbox1=\vbox{ \offinterlineskip\hrule
\halign{
&\vrule#&\strut\hfil\ #\ \hfil&\vrule#&\strut\hfil\ #\ \hfil&
\vrule#&\strut\hfil\ #\ \hfil&
\vrule#&\strut\hfil\ #\ \hfil&\vrule#&\strut\hfil\ #\ \hfil\cr
 height2mm&\omit&&\omit&&\omit&&\omit&&\omit&\cr 
&\hfil \hfil&&\hfil DFGS \hfil&&KLL&
&\hfil CGL\hfil&
&\hfil PY\hfil& \cr
 height1mm&\omit&&\omit&&\omit&&\omit&&\omit&\cr
\noalign{\hrule} 
height1mm&\omit&&\omit&&\omit&&\omit&&\omit&\cr
&$a_0^{(0)}$&&\vphantom{\Big|}$0.228\pm0.032$&&$0.224\pm0.013$&
&\hfil$0.220\pm0.005$ \hfil&&
$0.230\pm0.015$& \cr 
\noalign{\hrule}
height1mm&\omit&&\omit&&\omit&&\omit&&\omit&\cr
&$a_0^{(2)}$&&\vphantom{\Big|}$-0.0382\pm0.0038$&&$-0.0343\pm0.0036$&
&\hfil$-0.0444\pm0.0010$ \hfil&&
$-0.0480\pm0.0046$& \cr 
\noalign{\hrule}
height1mm&\omit&&\omit&&\omit&&\omit&&\omit&\cr
&$b_0^{(0)}$&&\vphantom{\Big|}&&$0.252\pm0.011$&
&\hfil$0.280\pm0.001$ \hfil&&
$0.312\pm0.014$& \cr 
\noalign{\hrule}
height1mm&\omit&&\omit&&\omit&&\omit&&\omit&\cr
&$b_0^{(2)}$&&\vphantom{\Big|}&&$-0.075\pm0.015$&
&\hfil$-0.080\pm0.001$ \hfil&&
$-0.090\pm0.006$& \cr 
\noalign{\hrule} 
height1mm&\omit&&\omit&&\omit&&\omit&&\omit&\cr
&$a_1$&&\vphantom{\Big|}&&$39.6\pm2.4$&
&\hfil${\displaystyle 37.9\pm0.5}$ \hfil&&
$38.4\pm0.8\quad(\times\,10^{-3})$& \cr 
\noalign{\hrule}
height1mm&\omit&&\omit&&\omit&&\omit&&\omit&\cr
&\vphantom{\Big|}$b_1$ 
\phantom{\big|}&&\phantom{\Big|}&&$2.83\pm0.67$&
&\hfil ${\displaystyle5.67\pm0.13}
\vphantom{\big|}$  
\hfil&&\hfil 
$4.75\pm0.16\quad(\times\,10^{-3})$& \cr
\noalign{\hrule} 
height1mm&\omit&&\omit&&\omit&&\omit&&\omit&\cr
&$a_2^{(0)}$&&\vphantom{\Big|}&
&\hfil  \hfil&&$17.5\pm0.3$&&$18.70\pm0.41$
$\quad(\times\,10^{-4})$& \cr
\noalign{\hrule} 
height1mm&\omit&&\omit&&\omit&&\omit&&\omit&\cr
&\vphantom{\Big|}$a_2^{(2)}$&&&
&\hfil  \hfil&&$1.70\pm0.13$&&$2.78\pm0.37$
$\quad(\times\,10^{-4})$& \cr
\noalign{\hrule}
height1mm&\omit&&\omit&&\omit&&\omit&&\omit&\cr
&$b_2^{(0)}$&&\vphantom{\Big|}&
&\hfil  \hfil&&$-3.55\pm0.14$&&$-4.16\pm0.30$
$\quad(\times\,10^{-4})$& \cr
\noalign{\hrule} 
height1mm&\omit&&\omit&&\omit&&\omit&&\omit&\cr
&\vphantom{\Big|}$b_2^{(2)}$&&&
&\hfil  \hfil&&$-3.26\pm0.12$&&$-3.89\pm0.28$
$\quad(\times\,10^{-4})$& \cr
\noalign{\hrule}
height1mm&\omit&&\omit&&\omit&&\omit&&\omit&\cr
&\vphantom{\Big|}$a_3$&&$ $&
&\hfil  \hfil&
&$5.6\pm0.2$&&$6.3\pm0.4$
$\quad(\times\,10^{-5})$& \cr
\noalign{\hrule}}
\vskip.05cm}
\centerline{\box1}
\bigskip
{\noindent\petit
Units of $M_\pi$. The numbers  in the CGL column are   as given by 
CGL in  Table~2 and elsewhere in their text. \hb
In PY, the values for the D, F waves parameters are from the 
Froissart--Gribov representation. 
The rest are from the fits, improved with dispersion relations, except for $a_1$ and
$b_1$ that have been taken   as in ref.~5. }
\medskip
\centerline{\sc Table~3}
\smallskip
\centerrule{5truecm}}
\box0
}\endinsert

\booksection{8. Comments on  $b_1$, on the scalar form factor 
of the pion\hb   and on chiral perturbation
theory}
\vskip-0.5truecm
\booksubsection{8.1. The $b_1$ parameter, a sum rule and chiral perturbation theory}

\noindent
The parameter $b_1$ is one for which the different determinations in 
Table~3 vary more; for example, the CGL 
number is more than $4\,\sigma$ away from the result of PY, which follows from a very robust 
evaluation using the pion form factor  in ref.~5. 
For this reason, we will look at it from two more points of view. 
First, we will write a sum rule for $b_1$ 
that depends little on the high energy behaviour; 
and, secondly, we will look at $b_1$ 
from the point of view of 
chiral perturbation theory.

To get the sum rule we write a dispersion relation for the quantity
$$\dfrac{\partial}{\partial s}\,\left(\dfrac{F^{(I_s=1)}(s,0)}{s-4M^2_\pi}\right),$$
which we evaluate at threshold. 
Taking into account that
$$\dfrac{\partial}{\partial s}\,\left(\dfrac{F^{(I_s=1)}(s,0)}{s-4M^2_\pi}
\right)_{s=4M^2_\pi}=\dfrac{3M_\pi}{2\pi}b_1,$$
we obtain a  fastly convergent relation for $b_1$:

$$\eqalign{M_\pi b_1=&\,
=\tfrac{2}{3}\int_{4M^2_\pi}^\infty\dd s\,\Bigg\{
\tfrac{1}{3}\left[\dfrac{1}{(s-4M^2_\pi)^3}-\dfrac{1}{s^3}\right]\imag F^{(I_t=0)}(s,0)
+\tfrac{1}{2}\left[\dfrac{1}{(s-4M^2_\pi)^3}+\dfrac{1}{s^3}\right]\imag F^{(I_t=1)}(s,0)\cr
-&\,\tfrac{5}{6}\left[\dfrac{1}{(s-4M^2_\pi)^3}-
\dfrac{1}{s^3}\right]\imag F^{(I_t=2)}(s,0) \Bigg\}.
\cr}
$$
Most of the contribution to $b_1$ comes from the S0 and P waves at low energy, while  
all other contributions (in particular, the Regge contributions) 
are substantially smaller than $10^{-3}$.  Adding all pieces we find
$$b_1=(4.99\pm0.21)\times10^{-3}\;M_{\pi}^{-5},
\equn{(8.1)}$$
a value reasonably compatible with what we  found  in  
(4.4), with a completely different method,
 $b_1=(4.55\pm0.21)\times10^{-3}\;M_{\pi}^{-5}$
 (because of correlations, the distance is actually 
$\sim1\,\sigma$). We can combine both and find a precise estimate,
$$b_1=(4.75\pm0.16)\times10^{-3}\;M_{\pi}^{-5}.
$$
which is in fact the number reported in Table~3.

We next turn to ch.p.t. 
To one loop we have\ref{17}
$$b_1=\dfrac{M^{-1}_\pi}{288\pi^3f^4_\pi}\left\{-\bar{l}_1+\bar{l}_2+\tfrac{97}{120}\right\}
$$
and $f_\pi$ is the pion decay constant. 
If we replace the values of the  $\bar{l}_i$ given in CGL 
we would find (in units with $M_\pi=1$)
$$b_1=(4.09\pm0.81)\times 10^{-3} \quad\hbox{[one loop; CGL constants  $\bar{l}_i$ ]}.
\equn{(8.2)}$$
In order to bring this into agreement with the CGL number, $b_1=(5.67\pm0.13)\times 10^{-3}$, 
we would need  corrections  (two loop and higher) of $\sim40\%$.

One may get an interesting relation for $b_1$ 
 by  eliminating the constants
$\bar{l}_1,\,\bar{l}_2$ between $b_1$ and $a_{0+},\,a_{00}$ defined by
$$a_{0+}=\tfrac{2}{3}[a_2^{(0)}-a_2^{(2)}],\quad a_{00}=\tfrac{2}{3}[a_2^{(0)}+2a_2^{(2)}].$$
We get 
$$b_1=\tfrac{5}{2}\left[3a_{0+}-a_{00}\right]+\left(\tfrac{97}{120}+\tfrac{1}{8}\right)
\dfrac{1}{288\pi^3f^4_\pi M_\pi}.
\equn{(8.3)}$$
Although this is valid only to one loop, 
 the large higher order (two loop) correction due to the 
factor $1/f^4_\pi$ in (8.1) cancels almost completely. 
Moreover, the r.h.s. is dominated by $a_{0+}$, which is known with great accuracy:\ref{5} 
one has $a_{0+}=(10.61\pm0.14)\times10^{-4}\,M_{\pi}^{-5}$. 
If we  take the values of CGL for  $a_{00}$ and $a_{0+}$, 
we  obtain, from (8.3),
$$b_1=(4.95\pm0.11)\times10^{-3}\quad{\rm CGL}.
\equn{(8.4)}$$
This again is separated by several standard deviations from the value given in CGL. 
Contrarily to this, if we 
 use the values in ref.~5 for the  $a_{00}$, $a_{0+}$, we find
$$b_1=(4.44\pm0.21)\times10^{-3}\quad{\rm PY}.
\equn{(8.5)}$$
This number agrees  with the  value  
 obtained with  completely different methods (Table~3), 
thereby  showing the  consistency of the 
PY  phase shifts, with themselves and with chiral perturbation theory.

\booksubsection{8.2. Chiral perturbation theory and the scalar form factor of the pion}

\noindent
Finally, we  
 compare the 
low energy $\pi\pi$ parameters 
with the results of chiral perturbation theory (ch.p.t.). 
We will consider the cases of CGL and our (PY) analyses. 
In both we take the value of the parameter $\bar{l}_3=2.9\pm2.4$
from the old calculation of Gasser and Leutwyler,\ref{17} 
although we will discuss it a bit more later.
This parameter has very little influence in the results. 
We include in the comparison the quadratic scalar
 radius of the pion, $\langle r^2_{{\rm S},\pi}\rangle$. 
For the CGL case we take it as given by the Donoghue, Gasser and Leutwyler calculation,\ref{18}
which CGL accept; 
for the PY case we take the determination obtained in ref.~19.
This will be now discussed.

Their results (from experimental analysis of the 
scalar pion form factor) for the pion scalar radius, $\langle r^2_{{\rm S},\pi}\rangle$, 
and of the ch.p.t. constant $\bar{l}_4$, 
related to it by
$$\langle r^2_{{\rm S},\pi}\rangle=\dfrac{3}{8\pi^2f^2_\pi}\Big\{\bar{l}_4-\tfrac{13}{12}\Big\},
$$
differs between the calculations of Donoghue et al.,\ref{18} 
who find
$\bar{l}_4=4.4\pm0.2$, and those in ref.~19, where the value
$\bar{l}_4=5.4\pm0.5$ is obtained. 
Recently, Ananthanarayan et al.\ref{20} have repeated the calculation of ref.~18 
and conclude that it is confirmed (and thus that
 the result of ref.~19 is rejected by
$\sim2\,\sigma$).  Unfortunately, the authors in ref.~20 still use the old parametrization of 
Hyams et al. (Eq.~(12a) and Table~1 in ref~11a) for the $K$ 
matrix for $\pi\pi$ and $\bar{K}K$ scattering to 
evaluate the scalar form factor of the pion, $F_S(s)$. 
This is quite incompatible (by more than a factor {\sl three}) with what one finds for the
inelasticity, 
$1-\eta_0^{(0)}$, by direct measurements of the $\pi\pi\to\bar{K}K$ cross section;\ref{11}
see\fig~8.
Moreover, and unlike in the estimate of ref.~19, the authors in ref.~20 
find a phase of the scalar form factor, $\delta(s)$, clearly below $\pi$ even for energies as 
high as $s=2\,\gev^2$ where from QCD estimates one expects it to be {\sl above} $\pi$. 
In fact,  by 
using methods like those for the vector form factor,\fnote{Note that he 
calculation is now less rigorous than what one has in the vector case. 
Details of the discussion on the 
scalar form factor of the pion, 
and the evaluation of $\bar{l}_4$, 
will be given in a forthcoming  future article.} one finds  
that the scalar form factor of the pion behaves, at large $s$, as 
$$F_S(s)\simeqsub_{s\to\infty}Cf^2_\pi\big[\bar{m}^2_u(s)+\bar{m}^2_d(s)]/s.
$$
 Unlike for the vector case, however, the constant $C$
 can now not be calculated. 
From this it follows that
$$\delta(s)\simeqsub_{s\to\infty}\pi\left\{1+\dfrac{2d_m}{\log s/\lambdav^2}\right\}.
\equn{(8.6)}$$
Here $\lambdav\simeq0.3\,\gev$ is the QCD  scale and $d_m=12/(33-2n_f)$
 is the anomalous dimension of the
mass.
Thus, and although part of the criticism of ref.~20 is correct in that 
the final state interaction theorem allows for a difference of $\pi$ in 
the phases of S0 wave and form factor, we still consider 
that the result of ref.~19 is more reliable than that of refs.~18,~20: 
at least it is not incompatible with data on $\pi\pi\to\bar{K}K$ scattering, 
and agrees 
at high energy
 with QCD expectations. 
Moreover, 
it is consistent with what one finds fitting the 
experimental low energy 
parameters, as will be shown below.

\midinsert{
\setbox0=\vbox{{\psfig{figure=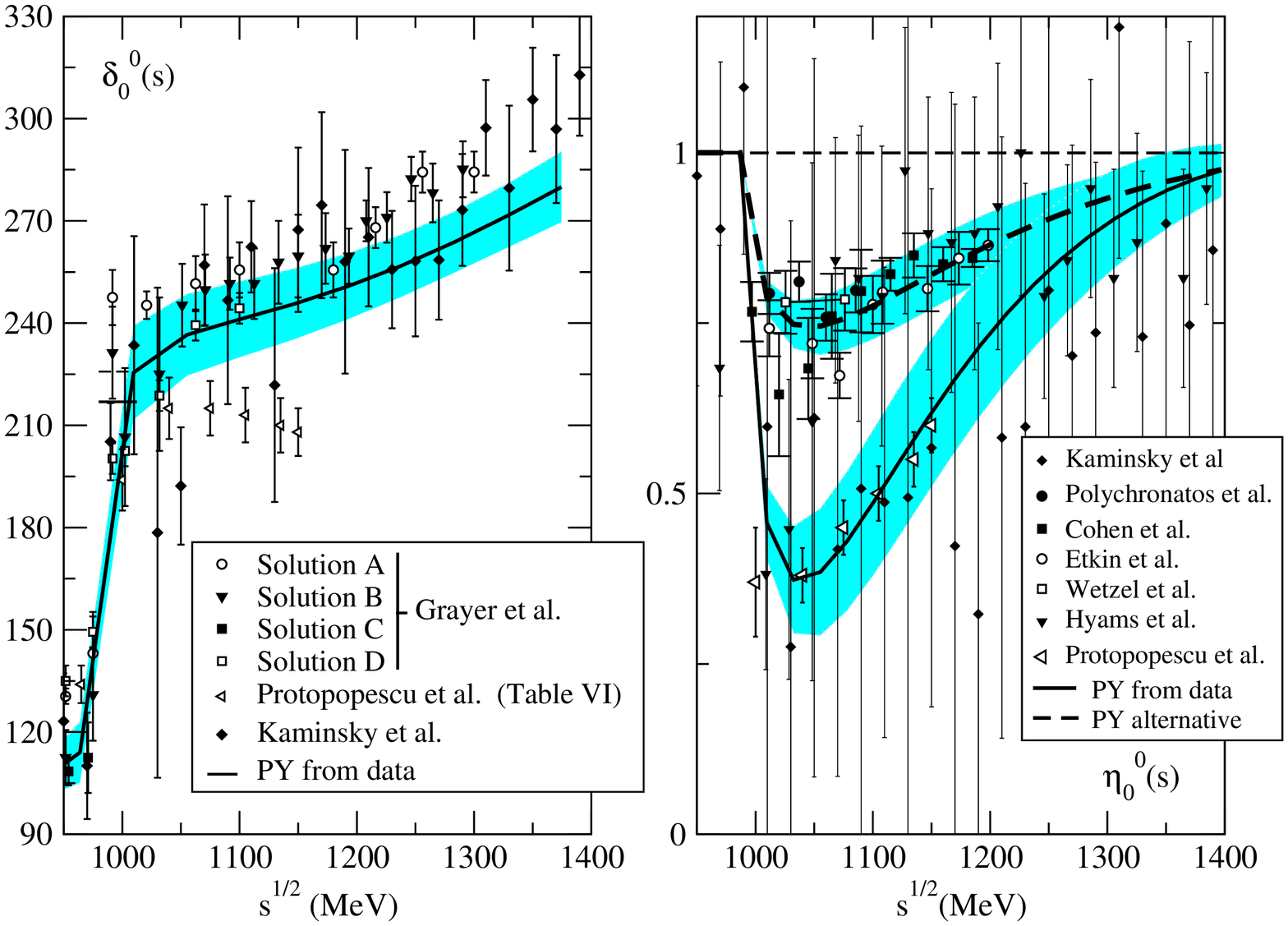,width=15.truecm,angle=-90}}} 
\setbox6=\vbox{\hsize 15truecm\captiontype\figurasc{Figure 8. }{Fit to the
 $I=0$, $S$-wave  inelasticity
 and phase shift  between 
 950 and 1400 \mev. 
Data  from refs.~10a,~10f. The 
difference between the determinations of $\eta_0^{(0)}$ from $\pi\pi\to\pi\pi$ 
(PY from data) and 
from $\pi\pi\to\bar{K}K$ measurements\ref{11} (PY alternative) is apparent here. 
Note that the value of $\eta_0^{(0)}$ hat follows from the parametrization of 
Hyams et al.\ref{11a}, that the authors in refs.~18,~20 use, lies at the lower 
part of the band denoted by ``PY from data" here.
}} 
\centerline{\tightboxit{\box0}}
\bigskip
\centerline{\box6}
\bigskip
}\endinsert

Another possibility would be to take the values for the chiral constant 
$\bar{l}_4$ from lattice calculations.\ref{21} 
The problem here is the existence of unknown, and potentially large 
theoretical and systematic errors. 
A last possibility is to obtain $\bar{l}_4$ by fitting the low energy parameters  
$a_2^{(I)}$, $b_1$, $b_0^{(I)}$, 
using for these last the ch.p.t results to one loop in ref.~17.
We give here, for ease of reference, the various numbers one finds:
$$\bar{l}_4=\cases{
4.4\pm0.2\quad \hbox{[refs.~18, 20]}\cr
5.4\pm0.5\quad \hbox{[ref.~19]}\cr
4.0\pm0.6\quad \hbox{[lattice calculation, ref.~21]}\cr
7.0\pm1.5\,{\rm exp.}\pm0.1\,{\rm l.o.}\quad \hbox{[fitting $a_2^{(I)}$, $b_1$,
$b_0^{(I)}$]}\cr }
\equn{(8.7)}$$
(for the explanation of the errors ``exp." and ``l.o.", see below).

The comparison of the low energy parameters and ch.p.t. for the CGL case 
is shown in Table~4; to get the one loop results there, 
we have taken the chiral lagrangian constants $\bar{l}_i$ as given 
by CGL themselves.
 
It is revealing that   almost all quantities require, to get 
agreement, rather large two loop corrections. 
This  is suggestive of bias in the input, a bias that 
we have shown before to exist with the help of dispersion relations.

\bigskip\setbox0=\vbox{\petit
\medskip
\setbox1=\vbox{\offinterlineskip\hrule
\halign{
&\vrule#&\strut\hfil#\hfil&\quad\vrule\quad#&\strut\quad#\quad&
\quad\vrule\quad#&\strut\quad#\quad&\quad\vrule#&\strut\quad#\quad&\quad\vrule#\cr
 height2mm&\omit&&\omit&&\omit&&\omit&\cr 
&\hfil\phantom{x} Quantity \hfil&&\hfil 2 loop and other  (CGL)\hfil&
&\hfil 1 loop ch.p.t. (CGL)&& \/l.o., ch.p.t.\/ \hfil& \cr
 height1mm&\omit&&\omit&&\omit&&\omit&\cr
\noalign{\hrule} 
height1mm&\omit&&\omit&&\omit&&\omit&\cr
&\phantom{\Big|}$a_0^{(0)}$&&\hfil $0.220\pm0.005$\hfil&&\hfil$0.204\pm0.006$&&$\phantom{-}0.157$ \hfil& \cr
\noalign{\hrule}
&\phantom{\Big|}$a_0^{(2)}$&&\hfil$-0.0444\pm0.0010$\hfil&&\hfil$-0.041\pm0.002$&&$-0.045$\hfil& \cr
\noalign{\hrule}
&\phantom{\Big|}$b_0^{(0)}$&&\hfil $0.280\pm0.010$\hfil&&\hfil $0.257\pm0.014$&&$\phantom{-}0.179$ 
\phantom{l}\hfil& \cr
\noalign{\hrule}
&\phantom{\Big|}$b_0^{(2)}$&&\hfil $-0.080\pm0.004$\hfil&&\hfil$-0.071\pm0.007$&&$-0.089$\hfil&\cr
\noalign{\hrule}
&\phantom{\Big|}$a_1$&&\hfil $(37.9\pm0.5)\times10^{-3}$\quad\hfil&&\hfil 
$(36.5\pm0.8)\times10^{-3}$&&$33.6\times10^{-3}$\hfil&\cr
\noalign{\hrule}
&\phantom{\Big|}$b_1$&&\hfil $(5.67\pm0.13)\times10^{-3}$\hfil&&\hfil 
$(4.09\pm0.81)\times10^{-3}$ \hfil&&\quad0&\cr
\noalign{\hrule}
&\phantom{\Big|}$a_2^{(0)}$&&\hfil $(17.5\pm0.3)\times10^{-4}$\hfil&&\hfil $(19.2\pm5.9)
\times10^{-4}$
\quad \hfil&&\quad0& \cr
\noalign{\hrule}
&\phantom{\Big|}$a_2^{(2)}$&&\hfil $(1.70\pm0.13)\times10^{-4}$\hfil&&\hfil $(4.4\pm1.6)
\times10^{-4}$
\quad\hfil&&\quad0& \cr
\noalign{\hrule}
&\phantom{\Big|}$\langle r^2_{{\rm S},\pi}\rangle$&
&\hfil $0.61\pm0.04\;{\rm fm}^2$\hfil&&\hfil $0.57\pm0.01\;{\rm fm}^2$\hfil&&\quad0&
\cr
\noalign{\hrule}}
\vskip.05cm}
\centerline{\box1}
\smallskip
\centerline{\sc Table~4}
\centerrule{6cm}
\medskip
\noindent{\petit Comparison of evaluations of low energy parameters 
 (according to CGL): leading order (``l.o.") and one loop 
(with the $\bar{l}_i$ parameters of CGL) 
and two loop 
chiral theory results and other results from CGL.}
\medskip}
\box0
\medskip

We then turn to out to what one finds from {\sl experiment}, 
i.e., using the low energy parameters from the 
 $\pi\pi$ amplitude determined here and in ref.~5, 
that we denote by (PY). 
We fix, from ref.~19, 
$$\bar{l}_4=5.4\pm0.5,
\equn{(8.7)}$$
and, from ref.~17,
$$\bar{l}_3=2.9\pm2.5.
\equn{(8.8)}$$

The D wave scattering lengths, and $b_1$, may be written in terms of the 
pion decay constant $f$ and the two 
chiral parameters $\bar{l}_1$, $\bar{l}_2$ only. 
For the first,  
$$\eqalign{
a_2^{(0)}=&\,\dfrac{M_\pi^{-1}}{1440\pi^3f^4}\big\{\bar{l}_1+4\bar{l}_2+
\tfrac{53}{8}\big\},\cr
a_2^{(2)}=&\,\dfrac{M_\pi^{-1}}{1440\pi^3f^4}\big\{\bar{l}_1+\bar{l}_2-
\tfrac{103}{40}\big\};\cr
\cr}
\equn{(8.9)}$$
the expression for $b_1$ was given in the previous 
Subsection.
We have an ambiguity here: we may take the pion decay constant as equal 
to the physical constant, 
$$f=f_\pi\simeq93.3\,\mev;$$  
or we can identify it with the decay constant in the chiral limit,
$$f=f_0=f_\pi\left\{1-\dfrac{M^2_\pi}{16\pi^2f^2_\pi}\bar{l}_4\right\}.
$$
We will in fact take the {\sl average} of both determinations, 
considering half the difference as an added error, called ``h.o.",  
which will be a (rough) estimate of the corresponding higher order 
ambiguity. 

From the scattering length combination $a_{0+}=\tfrac{2}{3}[a_2^{(0)}-a_2^{(2)}]$, which we 
determined very precisely (see Table~4 in ref.~5) we find
$$\bar{l}_2=5.48\pm0.06\,({\rm exp.})\pm0.61\,({\rm h.o.}).
\equn{(8.10a)}$$
From this and $a_2^{(0)}$,
$$\bar{l}_1=-0.76\pm0.37\,({\rm exp.})\pm0.29\,({\rm h.o.}).
\equn{(8.10b)}$$ 
 In Eqs.~(8.10), the first error is
that  due to the errors in the experimental $a_2^{(I)}$ and on $\bar{l}_3$, $\bar{l}_4$ 
from (8.7), (8.8), and the second error is that due to the difference between using the
two possibilities discussed for $f$, $f=f_0$ or $f=f_\pi$, in (8.9). 
Using this, we  obtain the results of Table~5.

\bigskip

\setbox0=\vbox{\petit
\medskip
\setbox1=\vbox{\offinterlineskip\hrule
\halign{
&\vrule#&\strut\hfil#\hfil&\quad\vrule\quad#&\strut\quad#\quad&
\quad\vrule\quad#&\strut\quad#\quad&\quad\vrule#\quad&\strut\quad#\quad&\quad\vrule#\cr
 height2mm&\omit&&\omit&&\omit&&\omit&\cr 
&\hfil\phantom{x} Quantity \hfil&&\hfil Exp. value (PY)\hfil&
&\hfil 1 loop ch.p.t. && \/l.o., ch.p.t.\/ \hfil& \cr
 height1mm&\omit&&\omit&&\omit&&\omit&\cr
\noalign{\hrule} 
height1mm&\omit&&\omit&&\omit&&\omit&\cr
&\phantom{\Big|}$a_0^{(0)}$&&\hfil
$0.230\pm0.015$\hfil&&\hfil$0.209\pm0.003\pm0.002$&&$\phantom{-}0.157$
\hfil& \cr
\noalign{\hrule}
&\phantom{\Big|}$a_0^{(2)}$&&\hfil$-0.048\pm0.005$\hfil&&\hfil$-0.042\pm0.001\pm0.001$&&
$-0.045$\hfil&
\cr
\noalign{\hrule}
&\phantom{\Big|}$b_0^{(0)}$&&\hfil $0.312\pm0.014$\hfil&&\hfil
$0.262\pm0.005\pm0.005$&&$\phantom{-}0.179$ 
\phantom{l}\hfil& \cr
\noalign{\hrule}
&\phantom{\Big|}$b_0^{(2)}$&&\hfil
$-0.090\pm0.006$\hfil&&\hfil$-0.073\pm0.001\pm0.003$&&$-0.089$\hfil&\cr
\noalign{\hrule}
&\phantom{\Big|}$a_1$&&\hfil $(38.4\pm0.8)\times10^{-3}$\quad\hfil&&\hfil 
$(37.6\pm0.6\pm0.6)\times10^{-3}$\hfil&&$33.6\times10^{-3}$\hfil&\cr
\noalign{\hrule}
&\phantom{\Big|}$b_1$&&\hfil $(4.75\pm0.16)\times10^{-3}$\hfil&&\hfil 
$(4.53\pm0.21\pm0.09)\times10^{-3}$ \hfil&&\quad0&\cr
\noalign{\hrule}
&\phantom{\Big|}$a_2^{(0)}$&&\hfil $(18.70\pm0.41)\times10^{-4}$\hfil&&\hfil\quad input
\quad \hfil&&\quad0& \cr
\noalign{\hrule}
&\phantom{\Big|}$a_2^{(2)}$&&\hfil $(2.78\pm0.37)\times10^{-4}$\hfil&&\hfil\quad input
\quad\hfil&&\quad0& \cr
\noalign{\hrule}
&\phantom{\Big|}$\langle r^2_{{\rm S},\pi}\rangle$&
&\hfil $0.77\pm0.07\;{\rm fm}^2$\hfil&&\hfil input\hfil&&\quad0&
\cr
\noalign{\hrule}}
\vskip.05cm}
\centerline{\box1}
\smallskip
\centerline{\sc Table~5}
\centerrule{6cm}
\medskip
\noindent{\petit Comparison of evaluations of low energy parameters 
from experiment (as determined in ref.~5), and from one loop 
chiral perturbation theory. Units of $M_\pi$ }
\medskip}
\box0
\medskip
 
For both  $a_0^{(0)}$ and $b_0^{(0)}$  already the one loop corrections 
(both for the CGL and PY evaluations) are quite  
large, so we expect also large two loop corrections, just by renormalization group
arguments.
In fact, a detailed estimate\ref{22} for $a_0^{(0)}$   
 gives 
$$\delta_{2\;{\rm loop}}\,a_0^{(0)}=0.017\pm0.002$$
which brings the ch.p.t. value of, e.g., PY, of the S0 scattering length to 
$$a_0^{(0)}=0.226\pm0.003\pm0.002\quad\hbox{[incl. two loop]},$$ 
almost on top of the experimental value, $0.230\pm0.015$. 
For almost all other quantities, the two loop corrections necessary 
to bring the theoretical result in agreement with experiment 
are smaller than what CGL find; only the effective range parameters 
$b_0^{(I)}$ do not have overlapping error bars.

We finish this article returning to the evaluation of $\bar{l}_3$, 
 $\bar{l}_4$. One can 
fix $\bar{l}_1$,  $\bar{l}_2$ from the scattering lengths $a_2^{(I)}$ as in Eqs.~(8.9), 
(8.10) 
and then fit  $\bar{l}_3$,  $\bar{l}_4$ from the all the  
other scattering lengths and effective range parameters. 
In this way one finds very poor \chidof's, which of course is 
just a reflection of the presence of higher order chiral corrections that we 
have not taken into account,\fnote{The error denoted by ``h.o." in 
our calculations only takes into
account the estimated influence of the higher order corrections for the 
$a_2^{(I)}$ on $\bar{l}_1$,  $\bar{l}_2$. } and the numbers
$$\bar{l}_4=7.1\pm0.7\,({\rm exp.)}\pm0.1\,({\rm h.o.});\quad
\bar{l}_3=3.5\pm10\,({\rm exp.)}\pm2.9\,({\rm h.o.}).
\equn{(8.11)}$$ 
The value of   $\bar{l}_4$ is almost identical to what we would obtain 
if fitting only $b_1$, $b_0^{(I)}$, 
$\bar{l}_4=7.0\pm0.7\pm0.1$, that we  had  already reported in 
(8.7). This  points clearly 
in  the direction of  
 what was found in ref.~19, and is well away from the 
results of refs.~18,~20. 
The value of 
$\bar{l}_3$ is compatible (within its huge error) with (8.8).

\vfill\eject
\booksection{Appendix. A comment on a recent paper by Caprini, Colangelo, Gasser and
Leutwyler}

\noindent
 In a recent paper Caprini, Colangelo, Gasser and
Leutwyler,\ref{23}  to be denoted by CCGL, review the work of Pel\'aez and the 
present author in ref.~3 
and conclude that, still, they consider the CGL solution 
consistent.
We have already discussed this in connection with 
the dispersion relations, from which one can conclude that 
the claims in CCGL are  
unjustified. 
We will here
examine only their contention   that the PY 
Reggeistics cannot be correct because it does not satisfy certain sum rules.
Of course this contention is meaningless since, as shown in the text (cf.~Fig.~4) 
the PY cross sections are perfectly compatible with
high energy ($s^{1/2}\geq1.42 \gev$) {\sl experimental} data, while the ACGL ones are not. 
However, there is perhaps some point in spending a few lines to examine this 
contention of CCGL.
 
CCGL evaluate the scattering length and effective range combinations 
$a_{0+}=\tfrac{2}{3}[a_2^{(0)}-a_2^{(2)}],\,a_{00}=\tfrac{2}{3}[a_2^{(0)}+2a_2^{(2)}]$ and 
the like combinations 
$b_{0+},\,b_{00}$ with the Froissart--Gribov  projection
 (as in refs.~3,~5) or with the Wanders's sum rules.  
They  define the differences $\Delta a_{0+}$, $\Delta b_{0+}$, \tdots, 
between the results of the two evaluations  and find, if using the Reggeistics of refs.~3,~4,
$$\eqalign{
\Delta a_{0+}=-0.11\times10^{-4},\quad \Delta a_{00}=0.034\times10^{-4},\cr
\Delta b_{0+}=-0.13\times10^{-4},\quad \Delta b_{00}=0.090\times10^{-4},\cr
}
\equn{(A.1)}$$
in units with $M_\pi=1$; cf. \sect~9 in CCGL.
From the fact that these numbers are not zero (as 
crossing symmetry would imply), CCGL conclude that 
the Reggeistics in refs.~3,~4  is not in ``equilibrium" with 
the low energy scattering amplitude.

However, errors should be taken into account, 
something that CCGL apparently forget. 
If we include only those errors following  from the  Froissart--Gribov 
representations in PY, and of these only those coming  from  D, F waves (since the 
S, and some of the P wave contributions cancel out) 
and from the Regge parameters, 
  we should replace the  relations (A.1) by 
$$\eqalign{
\Delta a_{0+}=(-0.11\pm0.09)\times10^{-4},\quad \Delta a_{00}=(0.03\pm0.09)\times10^{-4},\cr
\Delta b_{0+}=(-0.13\pm0.07)\times10^{-4},\quad \Delta b_{00}=(0.09\pm0.08)\times10^{-4}.\cr
}
\equn{(A.2)}$$
That is to say,  all the numbers are less than or around one standard deviation off zero,
 with one exception. This is $\Delta b_{0+}$, where
$-0.13$ is to be compared to $0.07$, in fact, 1.8 $\sigma$. 
 That among four  
quantities one is slightly off, by $1.8\,\sigma$ as  $\Delta b_{0+}$ is in our case, 
 is  statistically reasonable: you expect deviations of
that order about 
$20\%$ of the time and you get one in four. Not bad, 
particularly as this occurs for a quantity where the contribution of the 
slope of the 
$I_t=2$ exchange is large and (as stated several times in ref.~3) 
one cannot take very seriously discrepancies where the 
derivative of the isospin 2 exchange amplitude contributes substantially.
 This {\sl agreement} (A.2)  is the more remarkable because the 
low energy amplitude was obtained from fits 
to low energy $\pi\pi$ scattering: while the dominant Pomeron and $P'$
 Regge parameters come from 
a totally independent source, high energy  
$\pi N$ and $NN$ scattering, via factorization.  

Then, CCGL go on (still in their \sect~9) to claim that, for the ACGL, CGL  
asymptotics, the sum rules above hold to a remarkable degree of accuracy 
because they find, in this case,
$$\eqalign{
\Delta a_{0+}=-0.006\times10^{-4},\quad \Delta a_{00}=-0.009\times10^{-4},\cr
\Delta b_{0+}=-0.007\times10^{-4},\quad \Delta b_{00}=-0.03\times10^{-4}.\cr
}
\equn{(A.3)}$$
They consider this better than (A.1,2), and thus justifying their choice 
of Regge parameters, at least in an ``effective" manner.

The consistency of (A.3) is not too surprising:  ACGL {\sl fit} their Regge 
parameters by requiring consistency of  crossing sum rules. 
On the other hand, CCGL should not be so happy that the 
Eqs~(A.3) are fulfilled.  The relations (A.1-3) 
are such that the S waves, and partially the P waves, 
cancel out. 
So, they depend very crucially on the D waves. 
Since the ACGL amplitudes between 1.42 and 2 \gev\ are
 very distorted (as discussed in 
\sect~4 here and Appendix~C in ref.~5), in particular for the D2 wave 
(\fig~3), it follows that 
the Regge parameters they find must also be very distorted. 
So they are. 
Alternatively, and since the comparison with experiment of the Regge formulas of 
ACGL has revealed their inadequacy (Fig.~4), it follows that 
the lower energy piece ($s^{1/2}\leq2\gev$) of these authors 
has to be irrealistic, which it certainly is.

\vfill\eject
\brochuresection{Acknowledgments}
\noindent
We are grateful to CICYT, Spain, and to INTAS, for partial financial support.

\brochuresection{References}
\item{1 }{Ananthanarayan, B., Colangelo, G., Gasser, J.,  and Leutwyler, H.,
 {\sl Phys. Rep.}, {\bf
353}, 207,  (2001).}
\item{2 }{Colangelo, G., Gasser, J.,  and Leutwyler, H.,
 {\sl Nucl. Phys.} {\bf B603},  125, (2001).}
\item{3 }{Pel\'aez, J. R., and Yndur\'ain, F. J., {\sl Phys. Rev.} {\bf D68}, 074005 (2003).}
\item{4 }{Pel\'aez, J. R., and Yndur\'ain, F. J.,  {\sl Phys. Rev.} {\bf D69}, 114001 (2004).}
\item{5 }{Pel\'aez, J. R., and Yndur\'ain, F. J.,  FTUAM 04-14 (hep-ph/0411334).}
\item{6 }{de Troc\'oniz, J. F., and Yndur\'ain, F. J., {\sl Phys. Rev.},  {\bf D65}, 093001,
 (2002) 
and hep-ph/0402285. 
When quoting numbers, we will quote from this last paper.}
\item{7 }{Losty, M.~J., et al.  {\sl Nucl. Phys.}, {\bf B69}, 185 (1974); 
Hoogland, W., et al. 
{\sl Nucl. Phys.}, {\bf B126}, 109 (1977);  Cohen, D. et al., {\sl Phys. Rev.}
{\bf D7}, 661  (1973);
Durusoy,~N.~B., et al., {\sl Phys. Lett.} {\bf B45}, 517 (19730.}
\item{8 }{Rosselet, L., et al. {\sl Phys. Rev.} {\bf D15}, 574  (1977); 
Pislak, S.,  et al.  {\sl
Phys. Rev. Lett.}, {\bf 87}, 221801 (2001).}
\item{9 }{Protopopescu, S. D., et al., {\sl Phys Rev.} {\bf D7}, 1279, (1973).}
\item{10 }{Cern-Munich experiments: (a) Hyams, B., et al., {\sl Nucl. Phys.} {\bf B64}, 134,
(1973);  (b) Grayer, G., et al.,  {\sl Nucl. Phys.}  {\bf B75}, 189, (1974); 
(c)  Hyams, B., et al., {\sl Nucl. Phys.} {\bf B100}, 205 (1975). See also the
analysis of the  same experimental data in ; 
 (d) Au,~K.~L., Morgan,~D., and Pennington,~M.~R. {\sl Phys. Rev.} {\bf D35}, 1633 (1987).
(e) Estabrooks, P., and Martin, A. D., {\sl Nucl. Physics}, {\bf B79}, 301,  (1974). 
Including also polarized data: 
(f) Kami\'nski, R., Lesniak, L, and Rybicki, K., {\sl Z. Phys.} {\bf C74}, 79 (1997) and 
{\sl Eur. Phys. J. direct} {\bf C4}, 4 (2002).}
\item{11 }{$\pi\pi\to\bar{K}K$ scattering: 
Wetzel,~W., et al., {\sl Nucl. Phys.} {\bf B115}, 208 (1976);
Polychromatos,~V.~A., et al.,  {\sl Phys. Rev.} {\bf D19}, 1317 (1979);
Cohen, ~D. et al., {\sl Phys. Rev.} {\bf D22}, 2595 (1980); 
Etkin,~E. et al.,  {\sl Phys. Rev.} {\bf D25}, 1786 (1982).}
\item{12 }{Palou, F. P., and Yndur\'ain, F. J., {\sl Nuovo Cimento}, {\bf 19A}, 245, 
 (1974); Palou,~F.~P., S\'anchez-G\'omez,~J.~L., and Yndur\'ain,~F.~J., 
{\sl Z. Phys.}, {\bf A274}, 161, (1975).}
\item{13 }{Pennington, M. R., {\sl Ann. Phys.} (N.Y.), {\bf 92}, 164, (1975).}
\item{14 }{Biswas, N. N., et al., {\sl Phys. Rev. Letters}, 
{\bf 18}, 273 (1967) [$\pi^-\pi^-$, $\pi^+\pi^-$ and $\pi^0\pi^-$];
 Cohen, D. et al., {\sl Phys. Rev.}
{\bf D7}, 661  (1973) [$\pi^-\pi^-$];
 Robertson, W. J.,
Walker, W. D., and Davis, J. L., {\sl Phys. Rev.} {\bf D7}, 2554  (1973)  [$\pi^+\pi^-$]; 
Hoogland, W., et al.  {\sl Nucl. Phys.}, {\bf B126}, 109 (1977) [$\pi^-\pi^-$];
Hanlon, J., et al,  {\sl Phys. Rev. Letters}, 
{\bf 37}, 967 (1976) [$\pi^+\pi^-$]; Abramowicz, H., et al. {\sl Nucl. Phys.}, 
{\bf B166}, 62 (1980) [$\pi^+\pi^-$]. These  references cover the region
between  1.35 and 16 \gev, and agree within errors in the regions where they overlap 
(with the exception of $\pi^-\pi^-$ below 2.3 \gev, see the discussion in ref.~4).}
\item{15 }{Martin, B. R., Morgan, D., and and Shaw, G. {\sl 
Pion-Pion Interactions in Particle Physics}, Academic Press, New~York (1976).}
\item{16 }{Descotes, S., Fuchs, N. H.,  Girlanda, L., and   Stern, J., {Eur. Phys. J. C}, 
{\bf 24}, 469, (2002); Kami\'nski, R., Le\'sniak, L., and Loiseau, B.,
  {\sl Phys. Letters},
{\bf B551}, 241 (2003).}
\item{17 }{Gasser, J.,  and Leutwyler, H., {\sl Ann. Phys.} (N.Y.), {\bf 158}, 142  (1984).}
\item{18 }{Donoghue, J. F., Gasser, J.,  and Leutwyler, H., {\sl Nucl. Phys.},
 {\bf B343}, 341 (1990).}
\item{19 }{Yndur\'ain, F. J., {\sl Pys. Lett.} {\bf B578}, 99 (2004) and
 (E), {\bf B586}, 439 (2004).}
\item{20 }{Ananthanarayan, B., et al. IISc-CHEP-12/04 (hep-ph/0409222).}
\item{21 }{Aubin, C., et al. MILC Collaboration (hep-lat/0407028).}
\item{22 }{Colangelo, G., {\sl Phys. Letters} {\bf B350}, 85  (1995) and 
(E) {\bf B361}, 234 (1995); Bijnens, J.,  et al., {\sl Phys. Letters} {\bf B374}, 210 
 (1996).}
\item{23 }{Caprini, I., Colangelo, G.,  
Gasser, J., and Leutwyler, H., {\sl Phys. Rev.} {\bf D68}, 074006 (2003).}

\bye